\documentclass[a4paper,12pt]{article}

\usepackage{amsmath}
\usepackage{amssymb}
\usepackage{slashed}
\usepackage{amsthm}
\usepackage{xcolor}
\usepackage{calc}

\numberwithin{equation}{section}
\def\appendix#1{\addtocounter{section}{1}\setcounter{equation}{0}
\renewcommand{\thesection}{\Alph{section}}
\section*{Appendix \thesection\protect\indent \parbox[t]{11.15cm}{#1}}
\addcontentsline{toc}{section}{Appendix \thesection\ \ \ #1}}

\usepackage{accents}
\newcommand{\dbtilde}[1]{\accentset{\approx}{#1}}

\def\ciD{{\buildrel{\circ} \over \nabla}}

\newcommand{\bea}{\begin{eqnarray}}
\newcommand{\eea}{\end{eqnarray}}

\begin{document}

\begin{titlepage}
\begin{center}

\vspace*{-1.0cm}

\hfill  DMUS-MP-22-03
\\
\vspace{2.0cm}

\renewcommand{\thefootnote}{\fnsymbol{footnote}}
{\Large{\bf Supersymmetric $dS_n$ solutions for $n \geq 5$ in $D=11$ supergravity}}
\vskip1cm
\vskip 1.3cm
D. Farotti and J. Gutowski
\vskip 1cm
{\small{\it
Department of Mathematics,
University of Surrey \\
Guildford, GU2 7XH, UK.}\\
\texttt{d.farotti@surrey.ac.uk, j.gutowski@surrey.ac.uk}}

\end{center}
\bigskip
\begin{center}
{\bf Abstract}
\end{center}
We determine the necessary and sufficient conditions for warped product $dS_n$ solutions, 
$5 \leq n \leq 10$, to preserve supersymmetry in $D=11$ supergravity, without assuming factorization of the Killing spinors. We prove that for $7 \leq n \leq 10$, all such solutions are flat, with vanishing 4-form. We also show that the only warped product
$dS_6$ solutions are either the maximally supersymmetric $AdS_7 \times S^4$ solution, or
$\mathbb{R}^{1,6} \times N_4$ where $N_4$ is hyperK\"ahler, with vanishing 4-form.
Supersymmetric warped product $dS_5$ solutions are then classified;
it is shown that all such solutions are generalized M5-brane configurations, for
which the transverse space is $\mathbb{R} \times N_4$, and
$N_4$ is a hyperK\"ahler manifold. If the 4-form is covariantly constant, then $N_4$ admits a hyperK\"ahler potential.

\end{titlepage}

\section{Introduction}

De Sitter geometry is of significant interest in the context of string cosmology, and in terms
of holography. In the case of string cosmology, there have been numerous different approaches for investigating possible viable models. It has been known for some time that 
there are strict no-go theorems which imply that compactifications from regular warped product de-Sitter solutions are excluded \cite{gbds, deWit:1986mwo, Maldacena:2000mw}. Furthermore, additional phenomenological requirements, such as requiring the de-Sitter minima to be stable, and that the resulting models are compatible with slow roll inflation, also need to be satisfied \cite{Hertzberg:2007wc, Shiu:2011zt}. Taking such requirements into account, there has been significant progress in understanding how to construct viable models in string cosmology, including systematic analysis of viable models from the perspective of ${\mathcal{N}}=1, D=4$ supergravity \cite{Kallosh:2014oja}, as well as the derivation of models from novel $G_2$ geometric structures in $D=11$ \cite{Cribiori:2019hrb, Gunaydin:2020ric} and
$SU(3)$ structures from type IIA supergravity \cite{Danielsson:2009ff, Danielsson:2010bc, Danielsson:2011au}. The latter SU(3) structures
were obtained by taking a supersymmetric class of IIA $AdS_4$ warped product solutions \cite{Caviezel:2008ik, Lust:2004ig, Koerber:2008rx} and modifying the ansatz in such a way as to break the supersymmetry and admit $dS_4$ solutions. Notable further interest in de Sitter geometry has arisen from 
considerations of holography and entropy. The original gauge-gravity duality
\cite{Maldacena:1997re} formulated in terms of a conformal field theory dual 
to string theory in an $AdS$ space has been generalized in numerous ways.
In terms of de Sitter space, several proposed holographic dualities have been developed and applied, including \cite{Hawking:2000da, Strominger:2001pn, Strominger:2001gp, Kames-King:2021etp}.

Motivated by the importance of de Sitter solutions in string theory, in this work we initiate a systematic classification of supersymmetric de Sitter solutions in $D=11$ supergravity. In particular, we classify all supersymmetric warped product $dS_n$ solutions for $5 \leq n \leq 10$, for which the 4-form flux is invariant with respect to the isometries of $dS_n$. We shall leave the classification of warped product $dS_{2,3,4}$ solutions to future work. As observed, there are strict no-go theorems which hold for such geometries. Hence, we shall not make any assumptions on smoothness of the warp factor or 4-form flux, nor do we assume that the internal manifold is smooth or compact; the analysis will be done entirely locally. In addition, we do not make any assumptions regarding factorization of the Killing spinors, as assuming such factorization may produce an erroneous counting of supersymmetries, as observed in \cite{Gran:2016zxk}. The results of this paper therefore extend the classification constructed in \cite{Gutowski:2014ova} for supersymmetric $AdS$ and flat warped-product solutions in $D=11$ supergravity to include warped product $dS_n$ solutions $5 \leq n \leq 10$.  Classifications of warped product $AdS$
and flat geometries have also been constructed for type II supergravity \cite{Beck:2014zda, Beck:2015hpa}, and for
warped product $AdS$ geometries in heterotic supergravity \cite{Beck:2015gqa}.

In order to carry out the analysis of the conditions on the geometry and fluxes obtained from the Killing Spinor Equations (KSEs), we shall utilize spinorial geometry techniques, \cite{Gillard:2004xq, SpinorialMtheory}. In this method, the Killing spinor is written as a differential multi-form, and gauge transformations are used to simplify the structure of this multi-form into one (or more) simple canonical forms. This enables the KSEs to be written in terms of a linear system in the fluxes and spin connection, which can then be solved to extract the geometric conditions and the expression for the 4-form flux.

The plan of this paper is as follows. In section 2 we summarize the conditions on the bosonic fields associated with such warped product solutions obtained from the field equations and the Bianchi identities. We also analyse the KSEs and produce some results common to all of the $dS_n$ backgrounds. In Section 3 we use this to show that all warped product $dS_n$ backgrounds
for $7 \leq n \leq 10$ are flat, with vanishing 4-form. In Section 4 we analyse the supersymmetric warped product $dS_6$ backgrounds, and show that these are either the maximally supersymmetric  $AdS_7 \times S^4$ solution, or
$\mathbb{R}^{1,6} \times N_4$ where $N_4$ is hyperK\"ahler, with vanishing 4-form. In Section 5, we analyse the supersymmetric warped product $dS_5$ solutions, and we prove that these are warped products $\mathbb{R}^{1,5} \times_w N_5$ where $N_5$ is conformal to $\mathbb{R} \times N_4$, and
$N_4$ is a hyperK\"ahler manifold. Such solutions are generalized M5-brane geometries, for
which the transverse space is $\mathbb{R} \times N_4$. It is also shown that
if the 4-form is covariantly constant, then $N_4$ admits a hyperK\"ahler
potential. Some brief conclusions are presented in Section 6. There are also
several Appendices. Appendix A contains a summary of curvature components associated with the warped product $dS$ solutions. Appendix B contains further details of how the KSEs are integrated up along the de-Sitter directions, and then reduced to a gravitino KSE on the internal manifold.
Some expressions for integrability conditions obtained from the KSEs are also presented. Appendices C and D contain more details of the analysis of the
linear system obtained from the KSEs for $dS_6$ and $dS_5$ backgrounds, respectively. Appendix E contains more detail on the analysis of the
$dS_5$ solutions in the special case when the 4-form is parallel.

\section{Bosonic field equations and KSE}

The bosonic field equations of $D=11$ supergravity \cite{Cremmer} consist of the Einstein equations
\begin{eqnarray}
R_{AB}=\frac{1}{12}F_{AC_1C_2C_3}F_B^{~~C_1C_2C_3}-\frac{1}{144}g_{AB}F^2
\label{Einstein11D}
\end{eqnarray}
and the gauge field equations
\begin{eqnarray}
d\star_{11}F-\frac{1}{2}F\wedge F=0
\label{4formEOMS}
\end{eqnarray}
where $A,B,\dots $ are 11-dimensional indices and $F$ is the 4-form flux. Moreover, the Bianchi identities read
\begin{eqnarray}
dF=0~.
\label{Bianchi}
\end{eqnarray}
The Killing spinor equations (KSE) of $D=11$ supergravity \cite{Cremmer} are 
\begin{eqnarray}
\nabla_A\epsilon=\bigg(\frac{1}{288}\Gamma_A^{~~B_1B_2B_3B_4}F_{B_1B_2B_3B_4}-\frac{1}{36}F_{AB_1B_2B_3}\Gamma^{B_1B_2B_3}\bigg)\epsilon
\nonumber \\
\label{KSE1}
\end{eqnarray}
where $\nabla$ is the $D=11$ Levi-Civita connection and $\epsilon$ is a Majorana spinor. \\
\indent 
In our work we study warped product $dS_n$ backgrounds in $D=11$ supergravity, with $n\ge 5$. The metric on the $D=11$ spacetime $M_{11}$ is 
\begin{eqnarray}
ds^2(M_{11})=A^2ds^2(dS_n)+ds^2(M_{11-n})
\label{EQ1}
\end{eqnarray}
where $A$ is a function of the co-ordinates of the Riemannian manifold $M_{11-n}$ and
\begin{eqnarray}
ds^2(dS_n)=\frac{1}{(1+\frac{k}{4}|x|^2)^2}\eta_{\mu\nu}dx^{\mu} dx^{\nu}~~~~~\mu,\nu=0,1,\dots n-1
\label{dSD}
\end{eqnarray}
is the metric tensor of $n$-dimensional de Sitter spacetime, with $|x|^2=\eta_{\mu\nu}x^{\mu}x^{\nu}$ and $k=\frac{1}{\ell^2}$. We  introduce on $M_{11}$ the co-frame
\begin{eqnarray}
\textbf{e}^{\mu}=\frac{A}{\mathcal{U}}dx^{\mu}~,~~~~~\textbf{e}^a=e^a_{~\alpha}(y)dy^{\alpha}
\label{viel1}
\end{eqnarray}
where $a=n,n+1,\dots ,\sharp$; $y^{\alpha}$ are the co-ordinates on $M_{11-n}$ and
\begin{eqnarray}
\mathcal{U}=1+\frac{k}{4}|x|^2~.
\label{R}
\end{eqnarray}
In terms of the co-frame \eqref{viel1}, the metric tensor \eqref{EQ1} reads
\begin{eqnarray}
ds^2(M_{11})=\eta_{\mu\nu}\textbf{e}^{\mu} \textbf{e}^{\nu}+ds^2(M_{11-n})
\label{11DM}
\end{eqnarray}
where
\begin{eqnarray}
ds^2(M_{11-n})=\delta_{ab}\textbf{e}^a \textbf{e}^b~.
\label{m11-n}
\end{eqnarray}
Requiring the 4-form $F$ to be invariant under the isometry group $O(n,1)$ of $dS_n$, we have
\begin{eqnarray}
F=X
\label{FdSn}
\end{eqnarray}
where $X$ is a 4-form on $M_{11-n}$. In the following we reduce the bosonic field equations \eqref{Einstein11D}, \eqref{4formEOMS}, the Bianchi identities \eqref{Bianchi} and the KSE \eqref{KSE1} on $M_{11-n}$. First, we decompose the Einstein equations \eqref{Einstein11D} on $M_{11-n}$, obtaining
\begin{eqnarray}
k(n-1)A^{-2}-A^{-1}\widetilde{\nabla}^a\widetilde{\nabla}_a A-(n-1)A^{-2}(\widetilde{\nabla} A)^2+\frac{1}{144}X^2=0
\nonumber \\
\label{Ein1}
\end{eqnarray}
\begin{eqnarray}
\widetilde{R}_{ab}=nA^{-1}\widetilde{\nabla}_a\widetilde{\nabla}_b A+\frac{1}{12}X_{ac_1c_2c_3}X_b^{~c_1c_2c_3}-\frac{1}{144}\delta_{ab}X^2
\label{Ein2}
\end{eqnarray}
where $\widetilde{\nabla}$ denotes the Levi-Civita connection on $M_{11-n}$ and $\widetilde{R}_{ab}$ is the Ricci tensor on $M_{11-n}$. Some details about the computation of the Ricci tensor of \eqref{11DM} are presented in Appendix A. Moreover, the gauge field equations \eqref{4formEOMS} can be decomposed as
\begin{eqnarray}
\tilde{d}(A^n\star_{11-n}X)=0
\label{maxwell}
\end{eqnarray}
where $\tilde{d}$ is the exterior derivative on $M_{11-n}$. Eventually, we reduce the Bianchi identities \eqref{Bianchi} on $M_{11-n}$, obtaining
\begin{eqnarray}
\tilde{d}X=0~.
\label{Bianchiid}
\end{eqnarray} 
Now let us perform the reduction of the KSE \eqref{KSE1} on $M_{11-n}$. The $A=\mu$ component of \eqref{KSE1} is given by
\begin{eqnarray}
\frac{\partial}{\partial x^{\mu}}\epsilon=\frac{1}{\mathcal{U}}\bigg(-\frac{k}{4}x^{\nu}\Gamma_{\nu\mu}+\Gamma_{\mu}\mathcal{C}\bigg )\epsilon
\label{mudesitD}
\end{eqnarray}
where{\footnote{If $\omega$ is a $p$-form on $M_{11-n}$ then $\slashed{\omega}=\omega_{a_1 \dots a_p} \Gamma^{a_1 \dots a_p}$}}
\begin{eqnarray}
\mathcal{C}=-\frac{1}{2}\widetilde{\slashed{\nabla}}A+\frac{A}{288}\slashed{X}~.
\label{CD}
\end{eqnarray}
Equation \eqref{mudesitD} implies a partial differential equation for $\epsilon$
\begin{eqnarray}
\frac{\partial}{\partial x^{\mu}}\frac{\partial}{\partial x^{\nu}}\epsilon+\frac{k}{4\mathcal{U}}\big(x_{\mu}\frac{\partial}{\partial x^{\nu}}\epsilon+x_{\nu}\frac{\partial}{\partial x^{\mu}}\epsilon\big)-\frac{k^2}{16\mathcal{U}^2}x_{\mu}x_{\nu}\epsilon+\frac{k}{4\mathcal{U}}\eta_{\mu\nu}\epsilon=0
\nonumber \\
\label{ksemu14}
\end{eqnarray}
whose solution is given by
\begin{eqnarray}
\epsilon=\mathcal{U}^{-1/2}\big(\psi+x^{\mu}\tau_{\mu}\big)
\label{ksemu21}
\end{eqnarray}
where $\psi$ and $\tau_{\mu}$ are Majorana spinors which depend only on the co-ordinates on $M_{11-n}$. Substituting \eqref{ksemu21} into \eqref{mudesitD}, we obtain the following conditions
\begin{eqnarray}
\tau_{\mu}=\Gamma_{\mu}\mathcal{C}\psi
\label{ksemu24}
\end{eqnarray}
and
\begin{eqnarray}
\bigg(\frac{k}{4}+\widehat{\mathcal{C}}\mathcal{C}\bigg)\psi=0
\label{ksemu26}
\end{eqnarray}
where $\widehat{\mathcal{C}}$ is defined by $\mathcal{C}\Gamma_{\mu}=\Gamma_{\mu}\widehat{\mathcal{C}}$, i.e.
\begin{eqnarray}
\widehat{\mathcal{C}}=\frac{1}{2}\widetilde{\slashed{\nabla}}A+\frac{A}{288}\slashed{X}~.
\label{CDx}
\end{eqnarray}
Substituting \eqref{CD} into \eqref{ksemu26}, we get\footnote{Notice that $\Gamma_{a_1a_2\dots a_8}=0$ for $n\ge 5$.}
\begin{eqnarray}
&&\bigg(\frac{k}{4}-\frac{1}{4}(\widetilde{\nabla}A)^2+\frac{A}{72}X^{ba_1a_2a_3}(\widetilde{\nabla}_bA)\Gamma_{a_1a_2a_3}-\frac{A^2}{1152}X_{a_1a_2b_1b_2}X^{b_1b_2a_3a_4}\Gamma^{a_1a_2}_{~~~~~a_3a_4}
\nonumber \\
&&+\frac{A^2}{3456}X^2\bigg)\psi=0~.
\label{integrsusp1}
\end{eqnarray}
Furthermore, inserting \eqref{ksemu24} into \eqref{ksemu21}, we obtain
\begin{eqnarray}
\epsilon=\mathcal{U}^{-1/2}\bigg(1+x^{\mu}\Gamma_{\mu}\mathcal{C}\bigg)\psi~.
\label{epsilonfinale}
\end{eqnarray}
The $A=a$ component of the KSE \eqref{KSE1} is given by
\begin{eqnarray}
\widetilde{\nabla}_a\epsilon=\bigg(\frac{1}{288}\Gamma_a^{~~b_1\dots b_4}X_{b_1\dots b_4}-\frac{1}{36}X_{ab_1b_2b_3}\Gamma^{b_1b_2b_3}\bigg)\epsilon~.
\nonumber \\
\label{adsn}
\end{eqnarray}
Substituting \eqref{epsilonfinale} into \eqref{adsn}, we find
\begin{eqnarray}
\widetilde{\nabla}_a\psi=\sigma_{a}\psi
\label{ksem9}
\end{eqnarray}
and
\begin{eqnarray}
\bigg(\widetilde{\nabla}_a\mathcal{C}+\mathcal{C}\sigma_{a}-\widehat{\sigma}_{a}\mathcal{C}\bigg)\psi=0
\label{ksem92}
\end{eqnarray}
where
\begin{eqnarray}
\sigma_{a}=\frac{1}{288}\Gamma_a^{~~b_1\dots b_4}X_{b_1\dots b_4}-\frac{1}{36}X_{ab_1b_2b_3}\Gamma^{b_1b_2b_3}
\label{sigman2}
\end{eqnarray}
and 
\begin{eqnarray}
\widehat{\sigma}_{a} = -\frac{1}{288}\Gamma_a^{~~b_1\dots b_4}X_{b_1\dots b_4}+\frac{1}{36}X_{ab_1b_2b_3}\Gamma^{b_1b_2b_3} \ .
\end{eqnarray}
\indent
Hence, the reduction of the KSEs \eqref{KSE1} on $M_{11-n}$ produces a gravitino KSE \eqref{ksem9} on $M_{11-n}$, supplemented by \eqref{ksemu26} and \eqref{ksem92}, which are both quadratic in $X$. Equations \eqref{ksemu26} and \eqref{ksem92} arise as integrability conditions of \eqref{ksem9}, implementing the bosonic field equations \eqref{Ein1}-\eqref{maxwell} and the Bianchi identities \eqref{Bianchiid}. Some details about the reduction of the KSE along $M_{11-n}$ and the integrability conditions of \eqref{ksem9} are presented in Appendix B. \\
\indent

\subsection{$dS_6$ backgrounds}

In the case of $dS_6$ backgrounds, we define the 1-form on $M_6$
\begin{eqnarray}
s=\star_5 X~.
\label{sdef}
\end{eqnarray}
The bosonic field equations \eqref{Ein1}-\eqref{maxwell} and the Bianchi identities \eqref{Bianchiid}  read
\begin{eqnarray}
5kA^{-2}-A^{-1}\widetilde{\nabla}^a\widetilde{\nabla}_a A-5A^{-2}(\widetilde{\nabla}A)^2+\frac{1}{6}s^2=0
\nonumber \\
\label{laplacianA}
\end{eqnarray}
\begin{eqnarray}
\widetilde{R}_{ab}=6A^{-1}\widetilde{\nabla}_a\widetilde{\nabla}_b A+\frac{1}{3}\delta_{ab}s^2-\frac{1}{2}s_as_b
\label{eisteinm5}
\end{eqnarray}
\begin{eqnarray}
\tilde{d}(A^6s)=0
\label{gaugeds6}
\end{eqnarray}
\begin{eqnarray}
\tilde{d}\star_5 s=0~.
\label{bianchids6}
\end{eqnarray}
Moreover, the KSE \eqref{ksem9} are given by
\begin{eqnarray}
\widetilde{\nabla}_a\psi=\bigg(\frac{1}{12}s_a+\frac{1}{6}s^b\Gamma_{ba}\bigg)\Gamma^{(5)}\psi
\nonumber \\
\label{kseds6}
\end{eqnarray}
where
\begin{eqnarray}
\Gamma^{(5)}=\frac{1}{5!}\epsilon_{a_1a_2\dots a_5}\Gamma^{a_1a_2\dots a_5}
\end{eqnarray}
is the highest rank Gamma matrix on $M_5$.

\subsection{$dS_5$ backgrounds}

For $dS_5$ backgrounds, we define the 2-form on $M_6$
\begin{eqnarray}
G=\star_6 X~.
\label{gdef}
\end{eqnarray}
The bosonic field equations \eqref{Ein1}-\eqref{maxwell} and the Bianchi identities \eqref{Bianchiid} yield
\begin{eqnarray}
4kA^{-2}-A^{-1}\widetilde{\nabla}^a\widetilde{\nabla}_a A-4A^{-2}(\widetilde{\nabla}A)^2+\frac{1}{12}G^2=0
\label{eins456}
\end{eqnarray}
\begin{eqnarray}
\widetilde{R}_{ab}=5A^{-1}\widetilde{\nabla}_a\widetilde{\nabla}_b A+\frac{1}{6}G^2\delta_{ab}-\frac{1}{2}G_{cb}G^c_{~a}
\label{eins457}
\end{eqnarray}
\begin{eqnarray}
\tilde{d}(A^5 G)=0
\label{gaugeds5}
\end{eqnarray}
\begin{eqnarray}
\tilde{d}\star_6 G=0~.
\label{bianchids5}
\end{eqnarray}
Furthermore, the KSE \eqref{ksem9} are given by
\begin{eqnarray}
\widetilde{\nabla}_a\psi=\bigg(-\frac{1}{12}G_{ab}\Gamma^b+\frac{1}{12}\Gamma_a^{~~bc}G_{bc}\bigg)\Gamma^{(6)}\psi
\label{killingG}
\end{eqnarray}
where
\begin{eqnarray}
\Gamma^{(6)}=\frac{1}{6!}\epsilon_{a_1a_2\dots a_6}\Gamma^{a_1a_2\dots a_6}
\end{eqnarray}
is the highest rank Gamma matrix on $M_6$.

We remark that it is straightfoward to count the number of supersymmetries preserved by 
the warped product $dS_5$ backgrounds. In particular, if $\psi$ satisfies
({\ref{killingG}}) then so does $\Gamma_{\mu \nu} \psi$. On taking a frame basis for the $dS_5$ directions given by (frame indices are chosen to be compatible with the spinorial geometry 
calculation in Section 5):
\begin{eqnarray}
\{ {\bf{e}}^\pm = {1 \over \sqrt{2}} ({\bf{e}}^5 \pm {\bf{e}}^0) \ , \quad {\bf{e}}^\sharp \ ,  \quad {\bf{e}}^4 \ , \quad {\bf{e}}^9 \}
\end{eqnarray}
one can without loss of generality assume that $({\ref{killingG}})$ admits a solution with positive lightcone chirality $\psi = \psi_+$ where $\Gamma_+ \psi_+=0$. It then follows that
\begin{eqnarray}
\{\psi_+, \Gamma_{\sharp 4} \psi_+, \Gamma_{\sharp 9} \psi_+, \Gamma_{49} \psi_+,
\Gamma_{-\sharp} \psi_+, \Gamma_{-4} \psi_+, \Gamma_{-9} \psi_+, \Gamma_{-\sharp 49} \psi_+ \}
\end{eqnarray}
all satisfy $({\ref{killingG}})$. These 8 spinors are linearly independent. Hence it follows that warped product $dS_5$ solutions preserve 
$N=8$, $N=16$, $N=24$ or $N=32$ supersymmetries. The $N=32$ supersymmetric solutions of
$D=11$ supergravity are fully classified \cite{FGP}, and are $\mathbb{R}^{1,10}$, $AdS_7 \times S^4$, $AdS_4 \times S^7$, and a maximally supersymmetric plane wave solution. The 4-forms of $AdS_4 \times S^7$ and the maximally supersymmetric plane wave solution are not compatible with the 4-form for the warped product $dS_n$ solutions for $n \geq 5$. As we shall demonstrate in Section 5, the maximally supersymmetric $N=32$ warped product $dS_5$ solutions correspond to $\mathbb{R}^{1,10}$ and $AdS_7 \times S^4$. We shall however concentrate on the analysis of the $N=8$ solutions in this paper.

\section{Warped Product $dS_n$ backgrounds ($7 \leq n \leq 10$)}

In this section we show that warped $dS_n$ backgrounds, with $7 \leq n \leq 10$, have vanishing 11-dimensional Riemann tensor and vanishing 4-form flux $F$.
Let us consider first $n\ge 8$. For such values of $n$ the 4-form $X$ vanishes identically, thus $F=0$. Hence, the KSE \eqref{ksem9} simplify to
\begin{eqnarray}
\widetilde{\nabla}_a\psi=0
\label{tredsalto}
\end{eqnarray}
Equation \eqref{tredsalto} implies that $M_{11-n}$ is Ricci flat, that is
\begin{eqnarray}
\widetilde{R}_{ab}=0
\label{cfg3}
\end{eqnarray}
hence
\begin{eqnarray}
\widetilde{R}_{abcd}=0~.
\label{cfg8}
\end{eqnarray}
Moreover, the integrability conditions \eqref{integrsusp1} reduce to
\begin{eqnarray}
\bigg(k-(\widetilde{\nabla}A)^2\bigg)\psi=0
\label{quattrodsalto}
\end{eqnarray}
thus
\begin{eqnarray}
k-(\widetilde{\nabla}A)^2=0~.
\label{fgrt4}
\end{eqnarray}
Furthermore, inserting \eqref{cfg3} into \eqref{Ein2} we find
\begin{eqnarray}
\widetilde{\nabla}_a\widetilde{\nabla}_b A=0~.
\label{cfg9}
\end{eqnarray}
Substituting \eqref{cfg8}, \eqref{fgrt4}, and \eqref{cfg9} into \eqref{Riemann}, we find that the 11-dimensional Riemann tensor vanishes entirely, hence $M_{11}\simeq \mathbb{R}^{1,10}$. \\
\indent
Now let us consider $n=7$. The integrability conditions \eqref{integrsusp1} yield
\begin{eqnarray}
\bigg(\frac{k}{4}-\frac{1}{4}(\widetilde{\nabla}A)^2+\frac{A}{72}f\epsilon^{ba_1a_2a_3}(\widetilde{\nabla}_bA)\Gamma_{a_1a_2a_3}+\frac{A^2}{144}w^2\bigg)\psi=0
\label{jan1}
\end{eqnarray}
where $w=\star_4 X$ is a function on $M_4$. Equation \eqref{jan1} implies that
\begin{eqnarray}
\frac{k}{4}-\frac{1}{4}(\widetilde{\nabla}A)^2+\frac{A^2}{144}w^2=0
\label{jan2}
\end{eqnarray}
and
\begin{eqnarray}
w\epsilon^{ba_1a_2a_3}(\widetilde{\nabla}_bA)\Gamma_{a_1a_2a_3}\psi=0~.
\label{jan3}
\end{eqnarray}
Equation \eqref{jan2} implies that $\widetilde{\nabla}A\ne 0$, since $k>0$. Hence locally we can adapt a frame such that
\begin{eqnarray}
\widetilde{\nabla}_7A\ne 0 ~,~\widetilde{\nabla}_8A=\widetilde{\nabla}_9A=\widetilde{\nabla}_{\sharp}A=0
\label{jan4}
\end{eqnarray}
where we have denoted by $7,8,9,\sharp$ the directions along $M_4$. Inserting \eqref{jan4} into \eqref{jan3}, it follows that $w=0$, thus
\begin{eqnarray}
X=0~.
\label{vanishX}
\end{eqnarray}
Using \eqref{vanishX}, the integrability conditions \eqref{jan2} reduce to
\begin{eqnarray}
k-(\widetilde{\nabla}A)^2=0
\label{jan6}
\end{eqnarray}
and the KSE \eqref{ksem9} simplify to $\widetilde{\nabla}_a\psi=0$, which in turn implies
\begin{eqnarray}
\widetilde{R}_{ab}=0
\label{jan5}
\end{eqnarray}
that is $M_4$ is Ricci flat. Moreover, inserting \eqref{vanishX} and \eqref{jan5} into \eqref{Ein2} we find
\begin{eqnarray}
\widetilde{\nabla}_a\widetilde{\nabla}_b A=0~.
\label{jan7}
\end{eqnarray}
Since $\tilde{d}A$ is non-zero and covariantly constant on $M_4$, then locally 
\begin{eqnarray}
M_4\simeq S^1\times M_3~.
\label{M4fiber}
\end{eqnarray}
Decomposing \eqref{jan5} along \eqref{M4fiber}, it follows that $M_3$ is Ricci flat, hence $M_3$ is flat, which in turn implies that $M_4$ has vanishing curvature tensor, i.e.
\begin{eqnarray}
\widetilde{R}_{abcd}=0~.
\label{jan8}
\end{eqnarray}
Inserting \eqref{jan6}, \eqref{jan7} and \eqref{jan8} into \eqref{Riemann}, we find that the 11-dimensional Riemann tensor vanishes entirely, thus $M_{11}\simeq\mathbb{R}^{1,10}$.

\section{Warped Product $dS_6$ backgrounds}

To begin the analysis of warped $dS_6$ backgrounds, consider the integrability conditions \eqref{integrsusp1}. Using \eqref{sdef}, we get
\begin{eqnarray}
\bigg(\frac{k}{4}-\frac{1}{4}(\widetilde{\nabla}A)^2+\frac{A}{72}X^{ba_1a_2a_3}(\widetilde{\nabla}_bA)\Gamma_{a_1a_2a_3}+\frac{A^2}{144}s^2\bigg)\psi=0~.
\label{intds61}
\end{eqnarray}
This implies that
\begin{eqnarray}
\frac{k}{4}-\frac{1}{4}(\widetilde{\nabla}A)^2+\frac{A^2}{144}s^2=0
\label{intds62}
\end{eqnarray}
and
\begin{eqnarray}
s_c\epsilon^{cba_1a_2a_3}(\widetilde{\nabla}_bA)\Gamma_{a_1a_2a_3}\psi=0~.
\label{intds63}
\end{eqnarray}
In particular, \eqref{intds62} implies that $\widetilde{\nabla}A\ne 0$, as $k>0$. Moreover, equation \eqref{intds63} is equivalent to
\begin{eqnarray}
\Gamma^{ab}s_a(\widetilde{\nabla}_bA)\psi=0~.
\label{intds64}
\end{eqnarray}
Since $\widetilde{\nabla}A\ne 0$, then without loss of generality, we can pointwise choose a frame such that
\begin{eqnarray}
\widetilde{\nabla}_6A\ne 0~,~~\widetilde{\nabla}_7A=\widetilde{\nabla}_8A=\widetilde{\nabla}_9A=\widetilde{\nabla}_{\sharp}A=0
\label{intds65}
\end{eqnarray}
where we have denoted by $6,7,8,9,\sharp$ the directions along $M_5$. Inserting \eqref{intds65} into \eqref{intds64}, we obtain
\begin{eqnarray}
s_i\Gamma^i\psi=0
\label{intds66}
\end{eqnarray}
where $a=(6,i)$, with $i=7,8,9,\sharp$. Equation \eqref{intds66} implies that $s_i=0$, hence pointwise there exists a function ${\cal{Y}}$ such that
\begin{eqnarray}
s={\cal{Y}}\tilde{d}A~.
\label{fdA}
\end{eqnarray}
Inserting \eqref{fdA} into the gauge field equations \eqref{gaugeds6}, we find
\begin{eqnarray}
\tilde{d}{\cal{Y}}\wedge \tilde{d}A=0~.
\label{gaugeds62}
\end{eqnarray}
Since $\tilde{d}A\ne 0$, then \eqref{gaugeds62} implies that ${\cal{Y}}={\cal{Y}}(A)$. The Bianchi identities \eqref{bianchids6}, the Einstein equations \eqref{laplacianA}-\eqref{eisteinm5} and the integrability conditions \eqref{intds62} read (${\cal{Y}}^{'}=\frac{d{\cal{Y}}}{dA}$)
\begin{eqnarray}
{\cal{Y}}^{'}(\widetilde{\nabla}A)^2+{\cal{Y}}\widetilde{\nabla}^a\widetilde{\nabla}_aA=0
\label{E1}
\end{eqnarray}
\begin{eqnarray}
5kA^{-2}-A^{-1}\widetilde{\nabla}^a\widetilde{\nabla}_aA-5A^{-2}(\widetilde{\nabla}A)^2+\frac{1}{6}{\cal{Y}}^2(\widetilde{\nabla}A)^2=0
\label{E2}
\end{eqnarray}
\begin{eqnarray}
\widetilde{R}_{ab}=6A^{-1}\widetilde{\nabla}_a\widetilde{\nabla}_bA+\frac{1}{3}\delta_{ab}{\cal{Y}}^2(\widetilde{\nabla}A)^2-\frac{1}{2}{\cal{Y}}^2\widetilde{\nabla}_aA\widetilde{\nabla}_bA
\label{E2BIS}
\end{eqnarray}
\begin{eqnarray}
k-(\widetilde{\nabla}A)^2+\frac{1}{36}{\cal{Y}}^2A^2(\widetilde{\nabla}A)^2=0
\label{E3}
\end{eqnarray}
respectively. In particular, equations \eqref{E1}, \eqref{E2}, \eqref{E3} yield
\begin{eqnarray}
\widetilde{\nabla}^a\widetilde{\nabla}_aA=-kA^{-1}+A^{-1}(\widetilde{\nabla}A)^2
\label{E4}
\end{eqnarray}
\begin{eqnarray}
(\widetilde{\nabla}A)^2=\frac{k}{1-\frac{A^2}{36}{\cal{Y}}^2}
\label{E5}
\end{eqnarray}
\begin{eqnarray}
{\cal{Y}}^{'}+\frac{A}{36}{\cal{Y}}^3=0~.
\label{E6}
\end{eqnarray}
There are two separate cases to consider, depending on whether ${\cal{Y}}$ vanishes or not. If ${\cal{Y}}=0$, then $s=0$ by means of \eqref{fdA}, and the KSE \eqref{kseds6} reduce to 
\begin{eqnarray}
\widetilde{\nabla}_a\psi=0~.
\label{KSEf0}
\end{eqnarray}
Equation \eqref{KSEf0} implies
\begin{eqnarray}
\widetilde{R}_{ab}=0~.
\label{RAB}
\end{eqnarray}
Using \eqref{RAB}, equation \eqref{E2BIS} implies
\begin{eqnarray}
\widetilde{\nabla}_a\widetilde{\nabla}_bA=0
\label{dada}
\end{eqnarray}
that is $dA$ is covariantly constant on $M_5$. Hence locally $M_5=S^1\times N_4$. Let us define the 1-form
\begin{eqnarray}
V=\frac{1}{\sqrt{k}}dA~.
\label{VVVV}
\end{eqnarray}
Without loss of generality, we can take $V=\textbf{e}^6$. Then
\begin{eqnarray}
ds^2(M_5)=(\textbf{e}^6)^2+ds^2(N_4)~.
\label{metric54}
\end{eqnarray}
Decomposing \eqref{RAB} along \eqref{metric54}, it follows that $N_4$ is Ricci flat.
Moreover, reducing the KSE \eqref{KSEf0} on \eqref{metric54}, we get
\begin{eqnarray}
\dbtilde{\nabla}_i \psi=0
\label{ksef=0}
\end{eqnarray}
where $\dbtilde{\nabla}$ is the Levi-Civita connection on $N_4$. Furthermore, using \eqref{VVVV} and \eqref{metric54}, the 11-dimensional tensor \eqref{EQ1} reads
\begin{eqnarray}
ds^2(M_{11})=ds^2(N_7)+ds^2(N_4)
\end{eqnarray}
where ($\mu,\nu=0,1,\dots 5$)
\begin{eqnarray}
ds^2(N_7)=\frac{A^2}{\big(1+\frac{k}{4}|x|^2\big)^2}\eta_{\mu\nu}dx^{\mu}dx^{\nu}+\frac{1}{k}dA^2~.
\label{cdcd}
\end{eqnarray}
The 7-dimensional manifold \eqref{cdcd} has vanishing Riemann tensor, hence $N_7\simeq \mathbb{R}^{1,6}$. \\
\indent
Now let us consider ${\cal{Y}}\ne 0$. Solving equation \eqref{E6}, we obtain
\begin{eqnarray}
{\cal{Y}}= \frac{\eta}{\sqrt{q+\frac{A^2}{36}}}
\label{f}
\end{eqnarray}
where $\eta^2=1$ and $q$ is a positive constant. Indeed, if $q=0$, then \eqref{E3} implies $k=0$, which is contradictory. Notice that if $\psi$ satisfies \eqref{kseds6}, then $\Gamma^{(5)}\psi$ satisfies \eqref{kseds6}. Moreover, $(\Gamma^{(5)})^2=1$. Hence, without loss of generality, we will assume that
\begin{eqnarray}
\Gamma^{(5)}\psi= \psi~.
\label{projG5}
\end{eqnarray}
Using \eqref{projG5} the KSE \eqref{kseds6} simplify to
\begin{eqnarray}
\widetilde{\nabla}_a\psi=\bigg(\frac{1}{12}s_a+\frac{1}{6}s^b\Gamma_{ba}\bigg)\psi~.
\nonumber \\
\label{kseds62}
\end{eqnarray}
Consider the following conformal transformation on $\psi$
\begin{eqnarray}
\psi=h(A)\Psi
\label{hath}
\end{eqnarray}
where $h(A)$ satisfies
\begin{eqnarray}
h^{-1}\frac{dh}{dA}=\frac{1}{12}{\cal{Y}}~.
\label{hath2}
\end{eqnarray}
Using \eqref{hath} and \eqref{hath2}, equation \eqref{kseds62} implies
\begin{eqnarray}
\widetilde{\nabla}_a\Psi=\frac{1}{6}s^b\Gamma_{ba}\Psi~.
\label{kseintermediate}
\end{eqnarray}
Now let us perform a conformal transformation on the co-frame
\begin{eqnarray}
\textbf{e}^a={\cal{T}}(A)\check{\textbf{e}}^a
\label{ea}
\end{eqnarray}
where ${\cal{T}}(A)$ satisfies
\begin{eqnarray}
{\cal{T}}^{-1}\frac{d{\cal{T}}}{dA}=- \frac{1}{3}{\cal{Y}}~.
\label{dF}
\end{eqnarray}
Implementing \eqref{ea} and \eqref{dF} in \eqref{kseintermediate}, we obtain
\begin{eqnarray}
\check{\nabla}_a\Psi=0
\label{Rabhat}
\end{eqnarray}
where $\check{\nabla}$ is the Levi-Civita connection in the conformal frame.
Equation \eqref{Rabhat} implies that
\begin{eqnarray}
\check{R}_{ab}=0
\label{checkR}
\end{eqnarray}
where $\check{R}_{ab}$ is the Ricci tensor in the conformal frame, which is given by
\begin{eqnarray}
\widetilde{R}_{ab}=\check{R}_{ab}-3{\cal{T}}^{-1}\widetilde{\nabla}_b\widetilde{\nabla}_a {\cal{T}}+\delta_{ab}\bigg(-{\cal{T}}^{-1}\widetilde{\nabla}^c\widetilde{\nabla}_c {\cal{T}}+4{\cal{T}}^{-2}(\widetilde{\nabla} {\cal{T}})^2\bigg)~.
\nonumber \\
\label{conformalold}
\end{eqnarray}
Using \eqref{checkR} and \eqref{conformalold}, the Einstein equations \eqref{E2BIS} read
\begin{eqnarray}
&&6A^{-1}\widetilde{\nabla}_a\widetilde{\nabla}_bA+\frac{1}{3}\delta_{ab}{\cal{Y}}^2(\widetilde{\nabla}A)^2-\frac{1}{2}{\cal{Y}}^2\widetilde{\nabla}_aA\widetilde{\nabla}_bA
\nonumber \\
&=&-3{\cal{T}}^{-1}\widetilde{\nabla}_b\widetilde{\nabla}_a {\cal{T}}+\delta_{ab}\bigg(-{\cal{T}}^{-1}\widetilde{\nabla}^c\widetilde{\nabla}_c {\cal{T}}+4{\cal{T}}^{-2}(\widetilde{\nabla} {\cal{T}})^2\bigg)~.
\nonumber \\
\label{conff2}
\end{eqnarray}
Using \eqref{E1}, \eqref{E6} and \eqref{dF}, equation \eqref{conff2} implies
\begin{eqnarray}
\widetilde{\nabla}_a s_b=0
\label{covconsts}
\end{eqnarray}
namely $s={\cal{Y}} dA$ is covariantly constant on $M_5$. Moreover, using \eqref{E5} and \eqref{f}, we find
\begin{eqnarray}
s^2=\frac{k}{q}~.
\end{eqnarray}
Furthermore, equation \eqref{E2BIS} can be rewritten as
\begin{eqnarray}
\widetilde{R}_{ab}=-\frac{1}{3}s_as_b+\frac{k}{3q}\delta_{ab}~.
\label{ricciri}
\end{eqnarray}
Define the 1-form
\begin{eqnarray}
V=\sqrt{\frac{q}{k}}s~.
\label{VV}
\end{eqnarray}
Without loss of generality we set $V=\textbf{e}^6$. Then, using \eqref{covconsts}, we have
\begin{eqnarray}
ds^2(M_5)=(\textbf{e}^6)^2+ds^2(N_4)~.
\label{m54}
\end{eqnarray}
Decomposing \eqref{ricciri} on \eqref{m54}, we get
\begin{eqnarray}
\dbtilde{R}_{ij}(N_4)=\frac{k}{3q}\delta_{ij}
\label{cdew}
\end{eqnarray}
where $i,j=7,8,9,\sharp$. Using \eqref{VV} and \eqref{sdef}, the 4-form flux $X$ reads
\begin{eqnarray}
X_{i_1i_2i_3i_4}=6c~\textrm{dvol}(N_4)
\label{XXXX}
\end{eqnarray}
where $c$ is a constant given by
\begin{eqnarray}
c=\frac{1}{6}\sqrt{\frac{k}{q}}~.
\label{cdef}
\end{eqnarray}
Moreover, using \eqref{VV} the KSE \eqref{kseintermediate} read
\begin{eqnarray}
\dbtilde{\nabla}_i \Psi=c\Gamma_6\Gamma_i\Psi
\label{nablapsi6}
\end{eqnarray}
where $\dbtilde{\nabla}$ is the Levi-Civita connection on $N_4$. Using \eqref{f}, \eqref{VV} and \eqref{m54}, the 11-dimensional metric tensor \eqref{EQ1} is given by
\begin{eqnarray}
ds^2(M_{11})=ds^2(N_7)+ds^2(N_4)
\label{dsm11cne0}
\end{eqnarray}
where 
\begin{eqnarray}
ds^2(N_7)=\frac{A^2}{(1+\frac{k}{4}|x|^2)^2}\eta_{\mu\nu}dx^{\mu}dx^{\nu}+\frac{q/k}{q+\frac{A^2}{36}}dA^2~.
\label{dxdx}
\end{eqnarray}
Although it appears that there are 
two independent constants $k$ and $q$ in this metric, the metric
\eqref{dxdx} can be rewritten solely in terms of $c$. Indeed, defining $x^{\mu}=\frac{1}{\sqrt{k}}y^{\mu}$ and $A=\sqrt{q}\tilde{A}$, equation \eqref{dxdx} implies
\begin{eqnarray}
ds^2(N_7)=\frac{\tilde{A}^2}{36c^2}\frac{\eta_{\mu\nu}dy^{\mu}dy^{\nu}}{\big(1+\frac{1}{4}|y|^2\big)^2}+\frac{1}{36c^2}\frac{d\tilde{A}^2}{1+\frac{\tilde{A}^2}{36}}~.
\label{dxdx2}
\end{eqnarray}
A direct computation shows the Riemann curvature tensor of \eqref{dxdx2} is
\begin{eqnarray}
R_{ABCD}(N_7)=-c^2\big(g_{AC}g_{BD}-g_{AD}g_{BC}\big)
\label{rm7}
\end{eqnarray}
where $x^A \in \{y^\mu, {\tilde{A}} \}$. Hence $N_7$ is locally isometric to $AdS_7$. \\
\indent 
In the following we investigate the conditions on $N_4$ dictated by supersymmetry in both cases ${\cal{Y}}=0$ and ${\cal{Y}}\ne 0$. The KSE are given by \eqref{ksef=0} and \eqref{nablapsi6}, respectively. We analyze both cases simultaneously by writing
\begin{eqnarray}
\dbtilde{\nabla}_i \Psi=c\Gamma_6\Gamma_i\Psi
\label{eqm4}
\end{eqnarray}
where $c$ is given by \eqref{cdef} if ${\cal{Y}}\ne 0$, and $c=0$ if ${\cal{Y}}=0$. Indeed, by setting $c=0$ in \eqref{eqm4}, and using \eqref{hath}, we recover \eqref{ksef=0}. Notice that \eqref{eqm4} implies that $||\Psi||^2$ is constant. Without loss of generality, we set
\begin{eqnarray}
||\Psi||^2=1~.
\label{norm1}
\end{eqnarray}
Equation \eqref{projG5} is equivalent to
\begin{eqnarray}
\Gamma_{6789\sharp}\Psi=\Psi~.
\label{projection}
\end{eqnarray}
Furthermore, if $\Psi$ is Killing, then $\Gamma_{05}\Psi$ is Killing. Moreover $[\Gamma_{05},\Gamma_{6789\sharp}]=0$, hence again without loss of
generality we will take
\begin{eqnarray}
\Gamma_{05}\Psi= \Psi~.
\label{maj3}
\end{eqnarray}
The spinor $\Psi$ is Majorana, thus \cite{SpinorialMtheory}
\begin{eqnarray}
\Psi &=&\alpha 1+\bar{\alpha}e_{1234}+we_{5}+\bar{w}e_{12345}+\rho^{J}e_{J5}-\frac{1}{3!}(\star_4\bar{\rho})^{Q_1Q_2Q_3}e_{Q_1Q_2Q_35}
\nonumber \\
&+&\tau^Je_J-\frac{1}{3!}(\star_4\bar{\tau})^{Q_1Q_2Q_3}e_{Q_1Q_2Q_3}+\frac{1}{2}(A^{IJ}-(\star_4\bar{A})^{IJ})e_{IJ}+\frac{1}{2}(B^{IJ}-(\star_4\bar{B})^{IJ})e_{IJ5}
\nonumber \\
\label{maj32}
\end{eqnarray}
where $I,J,Q_1,Q_2,Q_3=1,2,3,4$ and $\alpha$, $w$ $\rho$, $\tau^J$, $A^{IJ}$ and $B^{IJ}$ are complex valued functions. Also, given a Majorana spinor $\eta$, the Gamma matrices are defined as \cite{SpinorialMtheory}
\begin{eqnarray}
&&\Gamma_0\eta=-e_5\wedge\eta+i_{e_5}\eta
\nonumber \\
&&\Gamma_5\eta=e_5\wedge\eta+i_{e_5}\eta
\nonumber \\
&&\Gamma_j\eta=e_j\wedge\eta+i_{e_j}\eta
\nonumber \\
&&\Gamma_{5+j}\eta=i(e_j\wedge\eta-i_{e_j}\eta)~.
\label{gammaspinor}
\end{eqnarray}
Implementing the projections \eqref{projection} and \eqref{maj3}, the spinor \eqref{maj32} simplifies to
\begin{eqnarray}
\Psi=\sum_{i=1}^{8}a_i\sigma_i
\label{afterproj}
\end{eqnarray}
where $a_1,a_2,\dots a_8$ are real functions on $N_4$ and
\begin{eqnarray}
&&\sigma_1=i(1-e_{1234})~,~~\sigma_2=e_1+e_{234}~,~~~\sigma_3=e_2-e_{134}~,~~~\sigma_4=e_3+e_{124}
\nonumber \\
&&\sigma_5=e_4-e_{123}~,~~~\sigma_6=i(e_{12}+e_{34})~,~~~\sigma_7=i(e_{13}-e_{24})~,~~~\sigma_8=i(e_{14}+e_{23})
\nonumber \\
\label{psi}
\end{eqnarray}
Notice that the spinor \eqref{afterproj} has eight real degrees of freedom. The KSE \eqref{eqm4} are $Spin(4)$ gauge covariant, namely they are invariant under
\begin{eqnarray}
\Psi\to\Psi^{'}=e^{n^{ij}\Gamma_{ij}}\Psi
\end{eqnarray}
where $\Gamma_{ij}$ are the generators of $Spin(4)$ and $n^{ij}$ are functions on $N_4$. In the following, we exploit such $Spin(4)$ gauge freedom to simplify the spinor \eqref{afterproj}. To this end, let us start by defining the self-dual generators 
\begin{eqnarray}
&&\tilde{\Gamma}_{78}=\frac{1}{2}\big(\Gamma_{78}+\Gamma_{9\sharp})~,~~~\tilde{\Gamma}_{79}=\frac{1}{2}\big(\Gamma_{79}-\Gamma_{8\sharp})~,~~~\tilde{\Gamma}_{89}=\frac{1}{2}\big(\Gamma_{89}+\Gamma_{7\sharp})
\nonumber \\
\label{dual}
\end{eqnarray}
and the anti-self-dual generators
\begin{eqnarray}
&&\hat{\Gamma}_{78}=\frac{1}{2}\big(\Gamma_{78}-\Gamma_{9\sharp})~,~~~\hat{\Gamma}_{79}=\frac{1}{2}\big(\Gamma_{79}+\Gamma_{8\sharp})~,~~~\hat{\Gamma}_{89}=\frac{1}{2}\big(\Gamma_{89}-\Gamma_{7\sharp})~.
\nonumber \\
\label{antiselfdual}
\end{eqnarray}
Then $\mathfrak{spin}(4)\simeq\mathfrak{sp}(1)\oplus\mathfrak{sp}(1)$, where one copy of $\mathfrak{sp}(1)$ is generated by the self-dual generators \eqref{dual} and the other copy by the anti-self-dual generators \eqref{antiselfdual}. In the following, we denote the self-dual generators collectively by $\tilde{\Gamma}$ and the anti-self-dual generators by $\hat{\Gamma}$. It is straightforward to check that
\begin{eqnarray}
\ker(\tilde{\Gamma})=\textrm{Im}(\hat{\Gamma})=\textrm{span}\big\{\sigma_1-\sigma_2,\sigma_3+\sigma_6,\sigma_4+\sigma_7,\sigma_5+\sigma_8\big\} 
\label{kerim1}
\end{eqnarray}
and
\begin{eqnarray}
\textrm{Im}(\tilde{\Gamma})=\ker(\hat{\Gamma})=\textrm{span}\big\{\sigma_1+\sigma_2,\sigma_3-\sigma_6,\sigma_4-\sigma_7,\sigma_5-\sigma_8\big\}~.
\label{kerim2}
\end{eqnarray}
Now let us consider the maps
\begin{eqnarray}
&&\tilde{\Gamma}:\textrm{Im}(\tilde{\Gamma})\to\textrm{Im}(\tilde{\Gamma})~,~~\hat{\Gamma}:\textrm{Im}(\hat{\Gamma})\to\textrm{Im}(\hat{\Gamma})~.
\end{eqnarray}
Defining
\begin{eqnarray}
\hat{I}=\hat{\Gamma}_{78}~,~~\hat{J}=\hat{\Gamma}_{79}~,~~~\hat{K}=-\hat{\Gamma}_{89}
\end{eqnarray}
\begin{eqnarray}
\tilde{I}=\tilde{\Gamma}_{78}~,~~\tilde{J}=\tilde{\Gamma}_{79}~,~~~\tilde{K}=-\tilde{\Gamma}_{89}
\end{eqnarray}
it follows that $\{ {\hat{I}}, {\hat{J}}, {\hat{K}} \}$ and
$\{ {\tilde{I}}, {\tilde{J}}, {\tilde{K}} \}$ satisfy the algebra
of the imaginary unit quaternions. Define
\begin{eqnarray}
\hat{M}=e^{c_1\hat{I}+c_2\hat{J}+c_3\hat{K}}~,~~~\tilde{M}=e^{d_1\tilde{I}+d_2\tilde{J}+d_3\tilde{K}}
\label{tildem}
\end{eqnarray}
where $c_1,c_2,c_3$ and $d_1,d_2,d_3$ are real functions on $N_4$. Notice that $\hat{M},\tilde{M}\in Sp(1)$. The action of $Spin(4)=Sp(1)\otimes Sp(1)$ is defined by
\begin{eqnarray}
(\tilde{M},\hat{M})(\tilde{\Psi},\hat{\Psi})=(\tilde{M}\tilde{\Psi},\hat{M}\hat{\Psi})
\end{eqnarray}
where 
\begin{eqnarray}
\tilde{\Psi}=a_1(\sigma_1+\sigma_2)+a_2(\sigma_3-\sigma_6)+a_3(\sigma_4-\sigma_7)+a_4(\sigma_5-\sigma_8) 
\label{psipsi}
\end{eqnarray}
\begin{eqnarray}
\hat{\Psi}=b_1(\sigma_1-\sigma_2)+b_2(\sigma_3+\sigma_6)+b_3(\sigma_4+\sigma_7)+b_4(\sigma_5+\sigma_8)~.
\label{phiphi}
\end{eqnarray}
Using \eqref{tildem}, it is straightforward to show that one copy of $Sp(1)$ acts transitively on \eqref{psipsi} and the other copy acts transitively on \eqref{phiphi}. Hence, without loss of generality, we can take
\begin{eqnarray}
\hat{\Psi}=\alpha(\sigma_1-\sigma_2)~,~~~\tilde{\Psi}=\beta(\sigma_1+\sigma_2)
\label{interspin1}
\end{eqnarray}
where $\alpha$ and $\beta$ are real functions on $N_4$. Equation \eqref{interspin1} implies
\begin{eqnarray}
\Psi=\alpha \sigma_1+\beta \sigma_2~.
\label{psisemifin}
\end{eqnarray}
Implementing \eqref{norm1}, equation \eqref{psisemifin} implies
\begin{eqnarray}
\Psi=\frac{i}{\sqrt{2}} \cos\theta (1-e_{1234})+\frac{1}{\sqrt{2}}\sin\theta (e_1+e_{234})
\label{PSIPSI}
\end{eqnarray}
where $\theta$ is a real function on $N_4$. The linear system associated to the KSE \eqref{eqm4}, with $\Psi$ given by \eqref{PSIPSI}, is presented in Appendix C. If $c\ne 0$, the linear system  \eqref{i=7first}-\eqref{i=sharplast} implies that $\cos(2\theta)\ne 0$ and
\begin{eqnarray}
\textbf{e}_7\theta=\textbf{e}_8\theta=\textbf{e}_9\theta=0~,~~\textbf{e}_{\sharp}\theta=c~.
\label{spinsyst1}
\end{eqnarray}
Moreover, the non-vanishing components of the spin connection are given by
\begin{eqnarray}
\dbtilde{\Omega}_{7,7\sharp}=\dbtilde{\Omega}_{8,8\sharp}=\dbtilde{\Omega}_{9,9\sharp}=-2c\tan(2\theta)
\end{eqnarray}
\begin{eqnarray}
\dbtilde{\Omega}_{7,89}=-\dbtilde{\Omega}_{8,79}=\dbtilde{\Omega}_{9,78}=\frac{2c}{\cos(2\theta)}~.
\label{spinsystfin}
\end{eqnarray}
Using  \eqref{spinsyst1}-\eqref{spinsystfin}, we compute the Riemann tensor of $N_4$
\begin{eqnarray}
\dbtilde{R}_{ijkl}(N_4)=4c^2\big(g_{ik}g_{jl}-g_{il}g_{jk}\big)
\label{rm4}
\end{eqnarray}
hence $N_4$ is locally isometric to $S^4$. Equations \eqref{dsm11cne0}, \eqref{rm7} and \eqref{rm4} imply that $R(N_7)=-7\Lambda$ and $R(N_4)=8\Lambda$, where $\Lambda=R(M_{11})=6c^2$. Thus 
\begin{eqnarray}
M_{11}\simeq AdS_7(-7\Lambda)\times S^4(8\Lambda)~.
\label{fig1}
\end{eqnarray}
Also, equation \eqref{XXXX} is equivalent to
\begin{eqnarray}
F=\sqrt{6\Lambda}~\textrm{dvol}(S^4)~.
\label{fig2}
\end{eqnarray}
Equations \eqref{fig1} and \eqref{fig2} correspond to the maximally supersymmetric $AdS_7 \times S^4$ solution found in \cite{pvnt}. If $c=0$, the linear system \eqref{i=7first}-\eqref{i=sharplast} implies that $\theta$ is constant and
\begin{eqnarray}
\dbtilde{\Omega}_{i,jk}+\frac{1}{2}\dbtilde{\Omega}_{i,lm}\epsilon^{lm}_{~~~jk}=0~.
\label{hyperK}
\end{eqnarray}
In particular, \eqref{hyperK} implies that $N_4$ is a hyperK\"ahler manifold.\\
\indent 
To summarize, there are two different classes of warped $dS_6$ backgrounds, namely
\begin{itemize}
\item{Class I}
\begin{eqnarray}
M_{11}\simeq\mathbb{R}^{1,6}\times N_4~,~~~~~F=0
\end{eqnarray}
where $N_4$ is a 4-dimensional hyperK\"ahler manifold.
\item{Class II}
\begin{eqnarray}
M_{11}\simeq AdS_7(-7\Lambda)\times S^4(8\Lambda)~,~~~~F=\sqrt{6\Lambda}~\textrm{dvol}(S^4)~.
\end{eqnarray}
\end{itemize}
Notice that $D=11$ Minkowski spacetime with vanishing flux falls into class I.

\section{Warped Product $dS_5$ backgrounds}

Using \eqref{gdef},  the integrability conditions \eqref{integrsusp1} factorize as follows
\begin{eqnarray}
\bigg(\frac{\sqrt{k}}{2}\Gamma^{(6)}+\widehat{\mathcal{C}}\bigg)\bigg(\frac{\sqrt{k}}{2}\Gamma^{(6)}-\mathcal{C}\bigg)\psi=0 
\label{integr2226}
\end{eqnarray}
where
\begin{eqnarray}
\mathcal{C}=-\frac{1}{2}\widetilde{\slashed{\nabla}}A-\frac{A}{24}\slashed{G}\Gamma^{(6)} \ , \qquad
\widehat{\mathcal{C}} = \frac{1}{2}\widetilde{\slashed{\nabla}}A-\frac{A}{24}\slashed{G}\Gamma^{(6)} \ .
\label{C5}
\end{eqnarray}
Multiplying \eqref{integr2226} by $\Gamma^{(6)}$, we obtain
\begin{eqnarray}
\bigg(\frac{\sqrt{k}}{2}\Gamma^{(6)}+\mathcal{C}\bigg)\phi=0
\label{integrlambda}
\end{eqnarray}
where
\begin{eqnarray}
\phi=\bigg(-\frac{\sqrt{k}}{2}-\Gamma^{(6)}\mathcal{C}\bigg)\psi~.
\label{lambdaspinor}
\end{eqnarray}
Using \eqref{lambdaspinor} and \eqref{ksem92}, it follows that $\phi$ is Killing, i.e. it satisfies \eqref{ksem9}. If $\phi$ vanishes, then, by means of \eqref{lambdaspinor}
\begin{eqnarray}
\Gamma^{(6)}\mathcal{C}\psi=-\frac{\sqrt{k}}{2}\psi~.
\label{lambdazero}
\end{eqnarray}
If $\phi\ne0$, \eqref{integrlambda} implies
\begin{eqnarray}
\Gamma^{(6)}\mathcal{C}\phi=\frac{\sqrt{k}}{2}\phi
\end{eqnarray}
hence $\psi$ and $\phi$ satisfy the same KSE and algebraic condition, modulo $\sqrt{k}\to-\sqrt{k}$. Thus, without loss of generality, we can consider \eqref{killingG} and \eqref{lambdazero}. Using \eqref{lambdazero}, the KSE \eqref{killingG} read
\begin{eqnarray}
\widetilde{\nabla}_a\psi=\bigg(-A^{-1}\widetilde{\nabla}_aA-A^{-1}\widetilde{\nabla}_bA\Gamma_{a}^{~~b}-\frac{1}{4}G_{ab}\Gamma^b\Gamma^{(6)}-\sqrt{k}A^{-1}\Gamma_a\Gamma^{(6)}\bigg)\psi~.
\nonumber \\
\label{killingG20}
\end{eqnarray}
By performing a conformal transformation on the spinor
\begin{eqnarray}
\psi=A^{-1} \Psi
\end{eqnarray}
equation \eqref{killingG20} simplifies to
\begin{eqnarray}
\widetilde{\nabla}_a\Psi=\bigg(-A^{-1}\widetilde{\nabla}_bA\Gamma_{a}^{~~b}-\frac{1}{4}G_{ab}\Gamma^b\Gamma^{(6)}-\sqrt{k}A^{-1}\Gamma_a\Gamma^{(6)}\bigg)\Psi~.
\nonumber \\
\label{killingG2}
\end{eqnarray}
Now consider a conformal transformation on the co-frame
\begin{eqnarray}
\textbf{e}^a=A^{-2} \check{\textbf{e}}^a
\label{conformal}
\end{eqnarray}
With this choice, \eqref{killingG2} is equivalent to
\begin{eqnarray}
\check{\nabla}_a\Psi=\bigg(-\frac{1}{4}A^{2}\check{G}_{ab}\Gamma^b-\sqrt{k}A^{-3}\Gamma_a\bigg)\Gamma^{(6)}\Psi
\label{main1}
\end{eqnarray}
where $\check{\nabla}$ is the Levi Civita connection in the conformal frame. Moreover, the algebraic condition \eqref{lambdazero} in the conformal frame reads
\begin{eqnarray}
\bigg(\frac{1}{2}A^{2}\check{\nabla}_aA \Gamma^a\Gamma^{(6)}+\frac{A^{5}}{24}\check{G}_{ab}\Gamma^{ab}+\frac{\sqrt{k}}{2}\bigg)\Psi=0~.
\label{main2}
\end{eqnarray}
It will be convenient to rewrite the bosonic field equations \eqref{eins456}-\eqref{gaugeds5} and the Bianchi identities \eqref{bianchids5} in the conformal frame \eqref{conformal}:
\begin{eqnarray}
4kA^{-2}-A^{3}\check{\nabla}^a\check{\nabla}_a A+4A^2(\check{\nabla} A)^2+\frac{1}{12}A^8 \check{G}^2=0
\label{bosonicconf3}
\end{eqnarray}
\begin{eqnarray}
\check{R}_{ab}&=&-3A^{-1}\check{\nabla}_ a \check{\nabla}_b A+12A^{-2}\check{\nabla}_ aA \check{\nabla}_ b A-\frac{1}{2}A^4\check{G}^c_{~a}\check{G}_{cb}
\nonumber \\
&+&\delta_{ab}\bigg(8A^{-2}(\check{\nabla} A)^2-2A^{-1}\check{\nabla}^c\check{\nabla}_c A+\frac{1}{6}A^4 \check{G}^2\bigg)
\label{bosonicconf4}
\end{eqnarray}
\begin{eqnarray}
d(A^5G)=0
\label{bosonicconf1}
\end{eqnarray}
\begin{eqnarray}
d{\check{\star}}_6(A^{-4}G)=0~.
\label{bosonicconf2}
\end{eqnarray}

\subsection{Canonical form for the spinor $\Psi$}
In the following, we denote the $dS_5$ directions by $0,5,\sharp, 4, 9$ and the directions along $M_6$ by $1,2,3,6,7,8$. Introducing the spacetime co-frame ($p=1,2,3$ is a holomorphic $SU(3)$ index)
\begin{eqnarray}
\textbf{e}^{\pm}=\frac{1}{\sqrt{2}}(\textbf{e}^5 \pm \textbf{e}^0)~,~~~\textbf{e}^p=\frac{1}{\sqrt{2}}(\textbf{e}^p-i\textbf{e}^{p+5})~,~~~\textbf{e}^{\bar{p}}=\frac{1}{\sqrt{2}}(\textbf{e}^p+i\textbf{e}^{p+5})
\nonumber \\
\label{basisosc}
\end{eqnarray}
equations \eqref{11DM} and \eqref{m11-n} read
\begin{eqnarray}
ds^2(M_{11})=2\textbf{e}^+\textbf{e}^-+(\textbf{e}^{\sharp})^2+(\textbf{e}^4)^2+(\textbf{e}^9)^2+2\delta_{p\bar{q}}\textbf{e}^{p}\textbf{e}^{\bar{q}}~.
\end{eqnarray}
The Gamma matrices \eqref{gammaspinor} in the basis \eqref{basisosc} are given by
\begin{eqnarray}
\Gamma_+=\sqrt{2}i_{e_5}~,~~~\Gamma_-=\sqrt{2}e_5\wedge ~,~~~~\Gamma_p=\sqrt{2}e_p~,~~~\Gamma_{\bar{p}}=\sqrt{2}i_{e_{\bar{p}}}~.
\end{eqnarray}
Moreover
\begin{eqnarray}
\Gamma^{(6)}=\Gamma_{123678}~.
\end{eqnarray}
If $\Psi$ satisfies \eqref{main1}, then $\Gamma_{+-}\Psi$ satisfies \eqref{main1}. Also, $(\Gamma_{+-})^2=1$. Hence, without loss of generality, we take $\Psi=\Psi_+$, where $\Gamma_+\Psi_+=0$. The spinor $\Psi_+$ is a positive chirality Majorana spinor, thus \cite{SpinorialMtheory}
\begin{eqnarray}
\Psi_+=\alpha 1+\bar{\alpha}e_{1234}+\tau^Je_J-\frac{1}{3!}(\star_4\bar{\tau})^{N_1N_2N_3}e_{N_1N_2N_3}+\frac{1}{2}C^{IJ}e_{IJ}
\nonumber \\
\label{majoranaplus}
\end{eqnarray}
where $I,J,N_1,\dots=1,2,3,4$,
\begin{eqnarray}
C^{IJ}=A^{IJ}-(\star_4\bar{A})^{IJ}
\end{eqnarray}
and $\alpha$, $\tau^J$ and $A^{IJ}$ are complex functions on $M_6$. In the following we split $I=p,4$. Defining $\chi^p=C^{4p}$, equation \eqref{majoranaplus} implies
\begin{eqnarray}
\Psi_+&=&\alpha 1+\bar{\alpha}e_{1234}+\tau^p e_p+\tau^4 e_4+\frac{1}{2}\epsilon^{qr}_{~~p}\bar{\tau}^pe_{4qr}-\bar{\tau}^4e_{123}
\nonumber \\
&+&\chi^ pe_{4p}+\frac{1}{2}\epsilon^{pq}_{~~r}\bar{\chi}^r e_{pq}~.
\label{spon56}
\end{eqnarray}
Now let us utilize the $Spin(6)$ gauge freedom of the KSE \eqref{main1} to simplify \eqref{spon56}. First of all, let us consider a $\mathfrak{su}(3)$ transformation $S^{p\bar{q}}\Gamma_{p\bar{q}}$, where $S^{p\bar{q}}$ is traceless (1,1). Since the action of $SU(3)$ on $\mathbb{C}^3-\{0\}$ is transitive, then without loss of generality we can set $\chi^2=\chi^3=0$ in \eqref{spon56}, obtaining
\begin{eqnarray}
\Psi_+&=&\alpha 1+\bar{\alpha}e_{1234}+\tau^p e_p+\tau^4 e_4+\frac{1}{2}\epsilon^{qr}_{~~p}\bar{\tau}^pe_{4qr}-\bar{\tau}^4e_{123}
\nonumber \\
&-&\chi e_{14}+\bar{\chi}e_{23}
\label{spon57}
\end{eqnarray}
where $\chi=\chi^1$. Next, let us consider the generators
\begin{eqnarray}
&&T_1=\frac{1}{2}\big(\Gamma_{23}+\Gamma_{\bar{2}\bar{3}}\big)~,~~~T_2=\frac{i}{2}\big(\Gamma_{23}-\Gamma_{\bar{2}\bar{3}}\big)~,~~~T_3=T_1 T_2
\nonumber \\
\label{matrixT}
\end{eqnarray}
acting on
\begin{eqnarray}
&&v_1=1+e_{1234}~,~~~v_2=i(1-e_{1234})~,~~~v_3=-e_{14}+e_{23}~,~~~v_4=i(e_{14}+e_{23})~.
\nonumber \\
\label{VVVVVs}
\end{eqnarray}
$\{T_1,T_2,T_3\}$ generate a Lie algebra which is isomorphic to $\mathfrak{sp}(1)$. Moreover
\begin{eqnarray}
(T_1)^2=(T_2)^2=(T_3)^2=T_1T_2T_3=-1~.
\end{eqnarray}
Consider
\begin{eqnarray}
M=e^{c_1T_1+c_2T_2+c_3T_3}\in Sp(1)
\label{MSP1}
\end{eqnarray}
where $c_1,c_2,c_3$ are functions on $M_6$. The action of \eqref{MSP1} on \eqref{VVVVVs} is transitive, hence \eqref{spon57} simplifies to
\begin{eqnarray}
\Psi_+=g_1(1+e_{1234})+\tau^p e_p+\tau^4 e_4+\frac{1}{2}\epsilon^{qr}_{~~p}\bar{\tau}^pe_{4qr}-\bar{\tau}^4e_{123}
\label{psithisstage}
\end{eqnarray}
where $g_1$ is a real function on $M_6$. Next, we perform a $SU(3)$ transformation to set $\tau^2=\tau^3=0$ in \eqref{psithisstage}. Also, we denote $\tau^1=g_2$ and $\tau^4=g_3$, where $g_2$ and $g_3$ are complex functions on $M_6$. The spinor \eqref{psithisstage} then simplifies to
\begin{eqnarray}
\Psi_+=g_1(1+e_{1234})+g_2e_1+\bar{g}_2e_{234}+g_3 e_4-\bar{g}_3 e_{123}~.
\label{psithisstage2}
\end{eqnarray}
If $g_1\ne 0$, then, taking $g_2$ to be real, equation \eqref{psithisstage2} implies
\begin{eqnarray}
\Psi_+=g_1(1+e_{1234})+g_2(e_1+e_{234})+g_3 e_4-\bar{g}_3 e_{123}~.
\label{psithisstage3}
\end{eqnarray}
If $g_1=0$, then equation \eqref{psithisstage2} implies
\begin{eqnarray}
\Psi_+=g_2e_1+\bar{g}_2e_{234}+g_3 e_4-\bar{g}_3 e_{123}~.
\label{spinorg1=0}
\end{eqnarray}
The action of \eqref{matrixT} on 
\begin{eqnarray}
w_1=e_1+e_{234}~,~~~w_2=i(e_1-e_{234})~,~~~w_3=e_4-e_{123}~,~~~w_4=i(e_4+e_{123})
\nonumber \\
\label{ws}
\end{eqnarray} is transitive, hence the spinor \eqref{spinorg1=0} is gauge equivalent to 
\begin{eqnarray}
\Psi_+=h(e_4-e_{123})
\label{g1=0}
\end{eqnarray}
where $h$ is a real function on $M_6$.

\subsection{Analysis of the KSE \eqref{main1} and \eqref{main2}}

The linear system associated to \eqref{main1} and \eqref{main2}, with $\Psi$ given by \eqref{psithisstage3}, is presented in Appendix D. After some computation, we find that the linear system \eqref{1star}-\eqref{dopoMajalg6} does not admit a solution for the case
of the spinor in ({\ref{g1=0}}). For the case of the spinor ({\ref{psithisstage3}}),
the linear system implies that
\begin{eqnarray}
g_1=g_2=g~,~~~g_3=0 \ ,
\label{algebraiccon1}
\end{eqnarray}
and hence the spinor $\Psi_+$ is
\begin{eqnarray}
\Psi_+=g\big(1+e_{1234}+e_1+e_{234}\big)~.
\label{PSISU2}
\end{eqnarray}

The following conditions on the geometry and the 2-form flux (here we split the $\mathfrak{su}(3)$ indices as $\{1, P\}$ for $P=2,3$) are also obtained:
\begin{eqnarray}
&&\check{\Omega}_{1,PQ}=0~,~~\check{\Omega}_{\bar{1},PQ}=0~,~~\check{\Omega}_{1,P}^{~~~~P}=0
\label{singfin1}
\end{eqnarray}
\begin{eqnarray}
&&\check{\Omega}_{P,\bar{1}Q}=0~,~~\check{\Omega}_{P,1Q}=0
\end{eqnarray}
\begin{eqnarray}
\check{\nabla}_1 \log g+\check{\nabla}_{\bar{1}}\log g +\frac{1}{2}(\check{\Omega}_{1,1\bar{1}}-\check{\Omega}_{\bar{1},1\bar{1}})=0
\label{singfin2}
\end{eqnarray}
\begin{eqnarray}
\check{\nabla}_1 \log g -\check{\nabla}_{\bar{1}}\log g +\frac{1}{2}(\check{\Omega}_{1,1\bar{1}}+\check{\Omega}_{\bar{1},1\bar{1}})=0
\label{singfin3}
\end{eqnarray}
\begin{eqnarray}
\frac{i\sqrt{2}}{2}A^2\check{G}_{1\bar{1}}-\check{\Omega}_{1,1\bar{1}}+\check{\Omega}_{\bar{1},1\bar{1}}=0
\label{singfin4}
\end{eqnarray}
\begin{eqnarray}
-\check{\Omega}_{1,1\bar{1}}-\check{\Omega}_{\bar{1},1\bar{1}}+2i\sqrt{2k}A^{-3}=0
\label{singfin5}
\end{eqnarray}
\begin{eqnarray}
\check{\nabla}_P \log g -\check{\Omega}_{\bar{1},1P}=0
\label{vfin1}
\end{eqnarray}
\begin{eqnarray}
\frac{i\sqrt{2}}{4}A^2 \check{G}_{1P}+\check{\Omega}_{\bar{1},1P}=0
\label{vfin2}
\end{eqnarray}
\begin{eqnarray}
\check{\Omega}_{P,1\bar{1}}=0~,~~~~\check{\Omega}_{P,Q}^{~~~~~Q}=0
\end{eqnarray}
\begin{eqnarray}
\check{\Omega}_{P,MN}=0~,~~\check{\Omega}_{P,\bar{M}\bar{N}}=0
\end{eqnarray}
\begin{eqnarray}
\check{\Omega}_{1,\bar{1}P}-\check{\Omega}_{\bar{1},1P}=0
\label{vfin4}
\end{eqnarray}
\begin{eqnarray}
\check{\Omega}_{1,1P}+\check{\Omega}_{\bar{1},1P}=0
\label{vfin5}
\end{eqnarray}
\begin{eqnarray}
\check{\Omega}_{\bar{1},\bar{1}P}+\check{\Omega}_{\bar{1},1P}=0
\label{vfin6}
\end{eqnarray}
\begin{eqnarray}
\check{\Omega}_{P,\bar{Q}\bar{1}}+\check{\Omega}_{P,\bar{Q}1}=0
\label{rfin1}
\end{eqnarray}
\begin{eqnarray}
\check{\Omega}_{P,\bar{Q}1}+\check{\Omega}_{\bar{Q},P\bar{1}}+\frac{i\sqrt{2}}{2}A^2 \check{G}_{P\bar{Q}}=0
\label{rfin3}
\end{eqnarray}
\begin{eqnarray}
\check{\Omega}_{P,\bar{Q}1}-\check{\Omega}_{\bar{Q},P\bar{1}}+2i\sqrt{2k}\delta_{P\bar{Q}}A^{-3}=0
\label{rfin4}
\end{eqnarray}
\begin{eqnarray}
&&\check{G}_{PQ}=0~,~~~\check{G}_P^{~~P}=0~,~~~\check{G}_{1P}+\check{G}_{\bar{1}P}=0
\label{algebraiccond2}
\end{eqnarray}
\begin{eqnarray}
\frac{i\sqrt{2}}{2}\big(\check{\nabla}_1 \log A-\check{\nabla}_{\bar{1}}\log A\big)=\sqrt{k}A^{-3}
\label{algebraincond2bis}
\end{eqnarray}
\begin{eqnarray}
\frac{i\sqrt{2}}{2}\big(\check{\nabla}_1 \log A+\check{\nabla}_{\bar{1}}\log A\big)=\frac{A^2}{6}\check{G}_{1\bar{1}}
\label{algebraincond2tris}
\end{eqnarray}
\begin{eqnarray}
\frac{i\sqrt{2}}{2}\check{\nabla}_P \log A=\frac{A^2}{6}\check{G}_{1P}~.
\label{algebraiccond3}
\end{eqnarray}

It will be convenient to introduce the $SU(2)$-invariant 1-form bilinears
\begin{eqnarray}
V_a=\langle \Psi_+, \Gamma_a \Psi_+\rangle~,~~~W_a=\langle\Psi_+,\Gamma_a\Gamma^{(6)}\Psi_+\rangle
\label{W}
\end{eqnarray}
and the $SU(2)$-invariant 2-forms
\begin{eqnarray}
\omega=-i\delta_{P\bar{Q}}\check{{\bf{e}}}^{P}\wedge \check{{\bf{e}}}^{\bar{Q}}~,~~~~\chi=2\check{{\bf{e}}}^{2}\wedge \check{{\bf{e}}}^3~.
\end{eqnarray}
Implementing \eqref{PSISU2}, \eqref{W} imply
\begin{eqnarray}
V=2\sqrt{2}g^2(\check{{\bf{e}}}^1+\check{{\bf{e}}}^{\bar{1}})~,~~~W=2\sqrt{2}ig^2(\check{{\bf{e}}}^1-\check{{\bf{e}}}^{\bar{1}})~.
\end{eqnarray}
A detailed analysis of equations \eqref{singfin1}-\eqref{algebraiccond3} produces the following covariant expression for the 2-form flux $G$
\begin{eqnarray}
A^5G=-\frac{1}{4}d(A^{3}g^{-2}W)~.
\label{Gfinale}
\end{eqnarray}
Observe that \eqref{Gfinale} implies that $A^5G$ is an exact 2-form, hence the gauge field equations \eqref{bosonicconf1} are automatically satisfied. Furthermore, the set of conditions on the geometry arising from \eqref{singfin1}-\eqref{algebraiccond3} is the following
\begin{eqnarray}
\mathcal{L}_W\log A=-4\sqrt{k}g^2A^{-3}
\label{LWLOGA}
\end{eqnarray}
\begin{eqnarray}
W=-\frac{4}{\sqrt{k}}A^3 g^2 d\log (g^{-2}A^{3})
\label{dlogafinale}
\end{eqnarray}
\begin{eqnarray}
dV=-2d\log (gA^{-3})\wedge V
\label{dVfinale}
\end{eqnarray}
\begin{eqnarray}
d\omega=4 d\log (g^{-2}A^{3})\wedge\omega
\label{domegafinale}
\end{eqnarray}
\begin{eqnarray}
d\chi=4d\log (g^{-2}A^{3})\wedge \chi~.
\label{dchifinale}
\end{eqnarray}
Equations \eqref{LWLOGA}-\eqref{dchifinale} are equivalent to
\begin{eqnarray}
\mathcal{L}_W\log A=-4\sqrt{k}t^{-1}
\label{LWLOGAT}
\end{eqnarray}
\begin{eqnarray}
W=-\frac{4}{\sqrt{k}}g^4 dt
\label{Wdt}
\end{eqnarray}
\begin{eqnarray}
V=g^{-2}A^6 ds
\label{Vds}
\end{eqnarray}
\begin{eqnarray}
d\hat{\omega}=0~,~~d\hat{\chi}=0
\label{dhatomegachi}
\end{eqnarray}
where
\begin{eqnarray}
t=g^{-2}A^3
\label{tcoordinate}
\end{eqnarray}
and $\hat{\omega}$ and $\hat{\chi}$ are computed with respect to the co-frame
\begin{eqnarray}
\hat{{\bf{e}}}^P=t^{-2}\check{{\bf{e}}}^P~.
\label{conformalt}
\end{eqnarray}

\subsection{Introducing local co-ordinates}

Using the geometric conditions \eqref{singfin2}, \eqref{singfin3} and \eqref{vfin4}, it follows that
\begin{eqnarray}
[V,W]=0
\label{commVWVW}
\end{eqnarray}
where $V$ and $W$ are the vector fields dual to \eqref{W}. Equation \eqref{commVWVW} implies that we can introduce two local co-ordinates $x,y$ such that
\begin{eqnarray}
V=\frac{\partial}{\partial x}~,~~~W=\frac{\partial}{\partial y}~.
\label{VWvec}
\end{eqnarray}
The 1-forms dual to \eqref{VWvec} are
\begin{eqnarray}
V=16g^4\big(dx+\alpha\big)~,~~~W=16g^4\big(dy+\beta\big)
\label{VW1forms}
\end{eqnarray}
with
\begin{eqnarray}
\alpha=\alpha_i dz^i~,~~~\beta=\beta_i dz^i
\end{eqnarray}
where we have denoted the co-ordinates on $M_6$ by $x,y,z^1,z^2,z^3,z^4$. The metric tensor in the conformal frame \eqref{conformal} is given by 
\begin{eqnarray}
ds^2(\check{M}_6)&=&\delta_{ab}\check{\textbf{e}}^a\check{\textbf{e}}^b
\nonumber \\
&=&\frac{1}{16g^4}V\otimes V+\frac{1}{16g^4}W\otimes W+2\delta_{P\bar{Q}}\check{\textbf{e}}^P \check{\textbf{e}}^{\bar{Q}}~.
\label{checkm6}
\end{eqnarray}
Notice that
\begin{eqnarray}
\check{\textbf{e}}^P=e^P_{~i}dz^i~.
\label{EPDZ}
\end{eqnarray}
Equation \eqref{dlogafinale} implies that
\begin{eqnarray}
d(g^{-2}A^{-3}W)=0
\end{eqnarray}
that is
\begin{eqnarray}
dW-d\log(g^2A^3)\wedge W=0~.
\label{dWEQ}
\end{eqnarray}
Inserting \eqref{VW1forms} into \eqref{dWEQ}, \eqref{dVfinale} and \eqref{LWLOGA}, one obtains
\begin{eqnarray}
A=yf_1(x,z)~,~~~g=yf_3(x,z)~,~~(f_1)^3=-4\sqrt{k}(f_3)^2
\label{aux903}
\end{eqnarray}
\begin{eqnarray}
t=-4\sqrt{k}y
\label{tyrel}
\end{eqnarray}
\begin{eqnarray}
V=16g^2 y^2\big(y^{-2}g^2 dx+\partial_{z^i}Ldz^i\big)
\label{Vuseful}
\end{eqnarray}
\begin{eqnarray}
W=16y^4(f_3)^4\big(dy+y\theta\big)
\label{WJAN}
\end{eqnarray}
where $L=L(x,z)$, with $\partial_x L=y^{-2}g^2$ and $\theta$ is a closed 1-form which does not depend on $x$ and $y$. Notice that \eqref{Vuseful} is equivalent to
\begin{eqnarray}
V=16g^2 y^2 dL~.
\end{eqnarray}
Comparing \eqref{Wdt} and \eqref{Vds} with \eqref{aux903}-\eqref{WJAN}, we find
\begin{eqnarray}
\theta=0~,~~~y^{-2}g^2 dx+{\partial L \over \partial z^i} dz^i=kds
\end{eqnarray}
where $s=s(x,z)$. Notice that equation \eqref{EPDZ} still holds in the $\{y,s,z^{i}\}$ co-ordinates. Inserting \eqref{Wdt} and \eqref{Vds} into \eqref{checkm6}, we get
\begin{eqnarray}
ds^2(\check{M}_6)=16k^2y^4ds^2+16y^4(f_3)^4dy^2+256k^2y^5 ds^2(N_4)
\label{checkm62}
\end{eqnarray}
where
\begin{eqnarray}
ds^2(N_4)=2\delta_{P\bar{Q}}\hat{\textbf{e}}^{P}\hat{\textbf{e}}^{\bar{Q}}=h_{ij}(z)dz^idz^j
\end{eqnarray}
is a hyperK\"ahler manifold independent of $y$ and $s$ (or $x$ and $y$). The manifold $N_4$ is hyperK\"ahler as a consequence of \eqref{dhatomegachi}. In order to show that the metric on $N_4$ is independent of $x$ and $y$, we find that as the Lie derivative of the hyperK\"ahler forms
with respect to $V$ vanishes, this implies that
\begin{eqnarray}
\mathcal{L}_V \hat{{\bf{e}}}^P= {\mathfrak{A}}^P{}_Q {\hat{{\bf{e}}}}^Q
\end{eqnarray}
for ${\mathfrak{A}} \in {\mathfrak{su}}(2)$. An appropriate $SU(2)$ gauge transformation can be 
utilised, which leaves the spinor $\Psi$ invariant, in order to set ${\mathfrak{A}}=0$,
and in this gauge
\begin{eqnarray}
\mathcal{L}_V \hat{{\bf{e}}}^P=0~.
\label{LVCHECK}
\end{eqnarray}
Similarly, one can also without loss of generality choose a gauge in which
\begin{eqnarray}
\mathcal{L}_W \hat{{\bf{e}}}^P=0~.
\label{LWCHECK}
\end{eqnarray}
Hence, we can take, without loss of generality
\begin{eqnarray}
\mathcal{L}_{\frac{\partial}{\partial x}}\hat{{\bf{e}}}^P=0~,~~\mathcal{L}_{\frac{\partial}{\partial y}}\hat{{\bf{e}}}^P=0
\end{eqnarray}
and therefore the metric on $N_4$ is independent on $x$ and $y$.

Now we compute the metric in the original frame. Using \eqref{conformal}, it follows that
\begin{eqnarray}
ds^2(M_6)=A^{-4}ds^2(\check{M}_6)~.
\label{checkm63}
\end{eqnarray}
Implementing \eqref{checkm62} in \eqref{checkm63}, we get
\begin{eqnarray}
ds^2(M_6)=16k^2(f_1)^{-4}ds^2+\frac{1}{k}(f_1)^2dy^2+256k^2(f_1)^{-4}ds^2(N_4)~.
\label{checkm64}
\end{eqnarray}
Defining $f=-\frac{1}{4\sqrt{k}}f_1$ and using \eqref{tyrel}, equation \eqref{checkm64} implies
\begin{eqnarray}
ds^2(M_6)=\frac{1}{16}f^{-4}(s,z^i)ds^2+\frac{1}{k}f^2(s,z^i)dt^2+f^{-4}(s,z^i)ds^2(N_4)~.
\end{eqnarray}
In the following we rescale $s=4\hat{s}$ and then we drop the hat on $s$
\begin{eqnarray}
ds^2(M_6)=f^{-4}(s,z^i)ds^2+\frac{1}{k}f^2(s,z^i)dt^2+f^{-4}(s,z^i)ds^2(N_4)~.
\label{M6rescaleds}
\end{eqnarray}
Moreover, notice that the first equation of \eqref{aux903} reads
\begin{eqnarray}
A=t\cdot f(s,z^i)~.
\label{Afine}
\end{eqnarray}
Using \eqref{M6rescaleds} and \eqref{Afine}, the 11-dimensional metric  \eqref{EQ1} is given by ($\mu,\nu=0,1,\dots 4$)
\begin{eqnarray}
ds^2(M_{11})&=&t^2f^2(s,z^i)\frac{\eta_{\mu\nu}dx^{\mu}dx^{\nu}}{\big(1+\frac{k}{4}|x|^2\big)^2}+f^{-4}(s,z^i)ds^2+\frac{1}{k}f^2(s,z^i)dt^2
\nonumber \\
&+&f^{-4}(s,z^i)h_{ij}(z)dz^idz^j~.
\label{11Dtensor}
\end{eqnarray}
Next, let us consider the 2-form flux \eqref{Gfinale}. Using  \eqref{Wdt} and \eqref{tcoordinate}, we obtain
\begin{eqnarray}
G=\frac{6}{\sqrt{k}}t^{-1}dA\wedge dt~.
\label{Gaux44}
\end{eqnarray}
Inserting \eqref{Afine} into \eqref{Gaux44}, we get
\begin{eqnarray}
G=\frac{6}{\sqrt{k}}df\wedge dt~.
\label{Gcoordinate}
\end{eqnarray}
A straightforward computation shows that
\begin{itemize}

\item{}  the Bianchi identities \eqref{bosonicconf2} are automatically satisfied, provided the KSE \eqref{main1}, the algebraic condition \eqref{main2}, the gauge field equations \eqref{bosonicconf1} and the Einstein equations \eqref{bosonicconf3} are satisfied.

\item{} the Einstein equations \eqref{bosonicconf4} are automatically satisfied, provided \eqref{bosonicconf3}-\eqref{bosonicconf1}, \eqref{main1} and \eqref{main2} hold.

\end{itemize}
Using \eqref{M6rescaleds} and \eqref{Gcoordinate}, the Einstein equations \eqref{bosonicconf3} reduce to
\begin{eqnarray}
\bigg(\frac{\partial f}{\partial s}\bigg)^2+h^{ij}\frac{\partial f}{\partial z^i}\frac{\partial f}{\partial z^j}-\frac{f}{7}\bigg(\frac{\partial^2 f}{\partial s^2}+h^{ij}\ciD_i \ciD_j f\bigg)=0
\label{einsm4}
\end{eqnarray}
where $\ciD$ denotes the Levi Civita connection on $N_4$. 

Further simplification of the solution can then be performed by setting $f=H^{-{1 \over 6}}$. The metric ({\ref{11Dtensor}}) can be written as
\begin{eqnarray}
ds^2(M_{11}) = H^{-{1 \over 3}} ds^2({\mathbb{R}}^{1,5}) + H^{{2 \over 3}} ds^2 ({\mathbb{R}} \times N_4)
\end{eqnarray}
and the Einstein equation ({\ref{einsm4}}) simplifies to
\begin{eqnarray}
\Box_5 H =0
\end{eqnarray}
where $\Box_5$ denotes the Laplacian on ${\mathbb{R}} \times N_4$. The 4-form is given by
\begin{eqnarray}
F = \star_5 dH
\end{eqnarray}
where $\star_5$ denotes the Hodge dual on ${\mathbb{R}} \times N_4$, and we set
\begin{eqnarray}
{\rm dvol}(M_6) = {1 \over \sqrt{k}}
H^{3 \over 2} dt \wedge {\rm dvol}
({\mathbb{R}} \times N_4) \ .
\end{eqnarray}
This geometry corresponds to that of a
generalized M5-brane configuration, whose transverse space is ${\mathbb{R}} \times N_4$ \cite{Gauntlett:1997pk}.

\subsection{Solutions with Parallel 4-form}

A special class of solutions arises when we take the 4-form $F$ to be covariantly
constant with respect to the Levi-Civita connection of the $D=11$ solution. It is known that $AdS_7 \times S^4$ lies within this class of solutions. Here we
shall investigate the additional conditions obtained on the geometry when
\begin{eqnarray}
\nabla^{(11)} F=0
\label{covconstX}
\end{eqnarray}
where $\nabla^{(11)}$ is the Levi-Civita connection on $M_{11}$. Using the 11-dimensional spin connection \eqref{spinconnection2} and \eqref{Gaux44}, equation \eqref{covconstX} is equivalent to
\begin{eqnarray}
\widetilde{\nabla} G=0
\label{covconstG}
\end{eqnarray}
where $\widetilde{\nabla}$ denotes the Levi Civita connection on $M_6$. Implementing \eqref{M6rescaleds} and \eqref{Gcoordinate}, equation \eqref{covconstG} gives the following set of PDEs
\begin{eqnarray}
\frac{\partial^2f}{\partial s^2}+f^{-1}\bigg(\frac{\partial f}{\partial s}\bigg)^2-2f^{-1}h^{ij}\frac{\partial f}{\partial z^i}\frac{\partial f}{\partial z^j}=0
\label{pde1}
\end{eqnarray}
\begin{eqnarray}
\frac{\partial f}{\partial s\partial z^i}+3f^{-1}\frac{\partial f}{\partial s}\frac{\partial f}{\partial z^i}=0
\label{pde2}
\end{eqnarray}
\begin{eqnarray}
\ciD_i \ciD_j f+3f^{-1}\frac{\partial f}{\partial z^i}\frac{\partial f}{\partial z^j}-2f^{-1}h_{ij}\bigg(\bigg(\frac{\partial f}{\partial s}\bigg)^2+h^{kl}\frac{\partial f}{\partial z^k}\frac{\partial f}{\partial z^l}\bigg)=0~.
\label{pde3}
\end{eqnarray}
The Einstein equations \eqref{einsm4} are implied by \eqref{pde1}-\eqref{pde3}. Moreover, \eqref{covconstG} implies that
\begin{eqnarray}
G^2=c^2
\label{constG}
\end{eqnarray}
where $c$ is constant. Implementing \eqref{constG} into \eqref{pde1}-\eqref{pde3}, we find
\begin{eqnarray}
f(s,z^i)=\sqrt{2}\bigg(c^2s^2+P(z)\bigg)^{\frac{1}{4}}
\label{fszizi}
\end{eqnarray}
\begin{eqnarray}
(\ciD P)^2=4c^2P
\label{usefulPP}
\end{eqnarray}
\begin{eqnarray}
\ciD_i \ciD_j P=2c^2 h_{ij}~.
\label{usefulhij}
\end{eqnarray}
Some details about the derivation of \eqref{fszizi}-\eqref{usefulhij} are presented in Appendix E.

We remark that in the case for which $c=0$, corresponding to $G=0$, these conditions imply that $f$ is constant. For such solutions, the metric $({\ref{11Dtensor}})$ simplifies to that of $\mathbb{R}^{1,6} \times N_4$ where $N_4$ is a hyperK\"ahler 4-manifold. On taking $N_4=\mathbb{R}^4$ one recovers the maximally supersymmetric flat $\mathbb{R}^{1,10}$ solution.
Otherwise, if $c \neq 0$, \eqref{usefulhij} implies that $\frac{P}{2c^2}$ is a hyperK\"ahler potential for $N_4$, in accordance with Proposition 5.6 of \cite{Swann}.

\section{Conclusion}

In this work we have classified the warped product $dS_n$ solutions of
$D=11$ supergravity for $5 \leq n \leq 10$. We have found that:

\begin{itemize}
\item[(i)] For $7 \leq n \leq 10$, all such solutions are flat, with vanishing 4-form. 
\item[(ii)] In the case of warped product $dS_6$ solutions, these are either the maximally supersymmetric $AdS_7 \times S^4$ solution, or
$\mathbb{R}^{1,6} \times N_4$ where $N_4$ is hyperK\"ahler, with vanishing 4-form.
\item[(iii)] In the case of warped product $dS_5$ solutions,
we prove that all such solutions are warped products $\mathbb{R}^{1,5} \times_w N_5$, as given in ({\ref{11Dtensor}}), where $N_5$ is conformal to $\mathbb{R} \times N_4$, and
$N_4$ is a hyperK\"ahler manifold. The geometry corresponds to a generalized M5-brane configuration, which
is determined by the choice of the hyperK\"ahler manifold, and a harmonic function on $\mathbb{R} \times N_4$. We have also shown that if the 4-form $F$ is covariantly constant, then the hyperK\"ahler manifold $N_4$ admits a hyperK\"ahler potential, and demonstrated how this potential is related to the warp factor.
\end{itemize}

In analysing the warped product supersymmetric $dS_5$ solutions, we have shown that such solutions preserve $N=8k$ supersymmetries ($k=1,2,3,4$). Our classification has determined the necessary and sufficient conditions for $N=8$ supersymmetry; the $N=32$ solutions are flat space and $AdS_7 \times S^4$. It would be interesting to determine the additional conditions imposed on the warp factor and on the hyperK\"ahler manifold $N_4$ in
order for there to be $N=16$ or $N=24$ enhanced supersymmetry, extending to the de-Sitter case the analysis for $AdS_5$ solutions with enhanced supersymmetry \cite{Beck:2016lwk}.

\setcounter{section}{0}
\setcounter{subsection}{0}

\appendix{Computation of the Ricci tensor}
\label{Riccitensor}
In this Appendix, we outline the computation of the Ricci tensor of \eqref{11DM}. First of all, the non-vanishing components of the spin connection in the frame \eqref{viel1} are
\begin{eqnarray}
\Omega_{\mu,\nu\rho}&=&kA^{-1}x_{[\nu}\eta_{\rho]\mu}
\nonumber \\
\Omega_{\mu,a\nu}&=&-\eta_{\mu\nu}A^{-1}{\tilde{\nabla}}_a A
\nonumber \\
\Omega_{a,bc}&=&\widetilde{\Omega}_{a,bc}
\label{spinconnection2}
\end{eqnarray}
where $\widetilde{\Omega}_{a,bc}$ denotes the spin connection on $M_{11-n}$. Using \eqref{spinconnection2}, the non-vanishing components of the 11-dimensional Riemann tensor are 
\begin{eqnarray}
R_{\mu\nu\rho\sigma}&=&\big(\eta_{\mu\rho}\eta_{\nu\sigma}-\eta_{\mu\sigma}\eta_{\nu\rho}\big)A^{-2}\big(k-(\widetilde{\nabla}A)^2\big)
\nonumber \\
R_{a\mu b\nu}&=&-\eta_{\mu\nu}A^{-1}\widetilde{\nabla}_a\widetilde{\nabla}_bA
\nonumber \\
R_{abcd}&=&\widetilde{R}_{abcd}
\label{Riemann}
\end{eqnarray}
where $\widetilde{R}_{abcd}$ is the Riemann tensor on $M_{11-n}$.  Implementing \eqref{Riemann}, we compute the 11-dimensional Ricci tensor, whose non-vanishing components are 
\begin{eqnarray}
R_{\mu\nu}&=&\eta_{\mu\nu}\bigg(k(n-1)A^{-2}-A^{-1}\widetilde{\nabla}^a\widetilde{\nabla}_aA-(n-1)A^{-2}(\widetilde{\nabla}A)^2\bigg)
\nonumber \\
R_{ab}&=&\widetilde{R}_{ab}-nA^{-1}\widetilde{\nabla}_a\widetilde{\nabla}_b A
\label{Ricci}
\end{eqnarray}
where $\widetilde{R}_{ab}$ is the Ricci tensor on $M_{11-n}$.

\appendix{Reduction of the KSE and integrability conditions}

In this Appendix, we provide some details about equations \eqref{ksemu14}, \eqref{ksemu21}, as well as the integrability conditions of \eqref{ksem9}. In the following, we denote $\partial_{\mu}=\frac{\partial}{\partial x^{\mu}}$. Acting on \eqref{mudesitD} with $\partial_{\nu}$, we find
\begin{eqnarray}
\partial_{\mu}\partial_{\nu}\epsilon&=&-\frac{k}{2\mathcal{U}}x_{\mu}\partial_{\nu}\epsilon-\frac{k}{4\mathcal{U}}\Gamma_{\mu\nu}\epsilon-\frac{k}{4\mathcal{U}}x^{\sigma}\Gamma_{\sigma\nu}\partial_{\mu}\epsilon-\frac{k}{4\mathcal{U}^2}\bigg(x^{\lambda}\Gamma_{\nu\lambda}\Gamma_{\mu}\mathcal{C}\epsilon
\nonumber \\
&+&x_{\nu}\Gamma_{\mu}\mathcal{C}\epsilon-x_{\mu}\Gamma_{\nu}\mathcal{C}\epsilon\bigg)
-\frac{1}{\mathcal{U}^2}\Gamma_{\mu\nu}\widehat{\mathcal{C}}\mathcal{C}\epsilon+\frac{1}{\mathcal{U}^2}\eta_{\mu\nu}\widehat{\mathcal{C}}\mathcal{C}\epsilon~.
\nonumber \\
\label{ksemu8}
\end{eqnarray}
Taking the symmetric part of \eqref{ksemu8}, we obtain
\begin{eqnarray}
\partial_{\mu}\partial_{\nu}\epsilon+\frac{k}{4\mathcal{U}}(x_{\mu}\partial_{\nu}\epsilon+x_{\nu}\partial_{\mu}\epsilon\big)-\frac{k^2}{16\mathcal{U}^2}x_{\mu}x_{\nu}\epsilon+\frac{1}{\mathcal{U}^2}\eta_{\mu\nu}\bigg(\frac{k^2|x|^2}{16}-\widehat{\mathcal{C}}\mathcal{C}\bigg)\epsilon=0~.
\nonumber \\
\label{ksemu9}
\end{eqnarray}
Moreover, anti-symmetrizing \eqref{ksemu8}, we get
\begin{eqnarray}
\bigg(\frac{k}{4}+\widehat{\mathcal{C}}\mathcal{C}\bigg)\epsilon=0~.
\label{ksemu13}
\end{eqnarray}
Substituting \eqref{ksemu13} into \eqref{ksemu9}, we obtain \eqref{ksemu14}. In order to solve \eqref{ksemu14}, we set
\begin{eqnarray}
\epsilon={\cal{F}}(|x|^2)\eta
\label{epsilon}
\end{eqnarray}
where $\eta$ is a Majorana spinor and ${\cal{F}}$ is a function. Substituting \eqref{epsilon} into \eqref{ksemu14}, we obtain 
\begin{eqnarray}
&&\eta_{\mu\nu}\big(2{\cal{F}}^{'}+\frac{k{\cal{F}}}{4\mathcal{U}}\big)\eta+{\cal{F}}\partial_{\mu}\partial_{\nu}\eta+(x_{\mu}\partial_{\nu}\eta+x_{\nu}\partial_{\mu}\eta)\big(2{\cal{F}}^{'}+\frac{k{\cal{F}}}{4\mathcal{U}}\big)
\nonumber \\
&+&x_{\mu}x_{\nu}\big(4{\cal{F}}^{''}+\frac{k}{\mathcal{U}}{\cal{F}}^{'}-\frac{k^2}{16\mathcal{U}^2}{\cal{F}}\big)\eta=0
\label{ksemu17}
\end{eqnarray}
Enforcing the vanishing of the coefficient of the $(x_{\mu}\partial_{\nu}\eta+x_{\nu}\partial_{\mu}\eta)$ term in \eqref{ksemu17} implies
\begin{eqnarray}
2{\cal{F}}^{'}+\frac{k{\cal{F}}}{4\mathcal{U}}=0
\end{eqnarray}
which implies (the integration constant has been absorbed into ${\cal{F}}$)
\begin{eqnarray}
{\cal{F}}=\mathcal{U}^{-1/2}~.
\label{ksemu18}
\end{eqnarray}
Inserting \eqref{ksemu18} in \eqref{ksemu17} we obtain
\begin{eqnarray}
\partial_{\mu}\partial_{\nu}\eta=0
\label{ksemu19}
\end{eqnarray}
whose solution is given by
\begin{eqnarray}
\eta=\psi+x^{\mu}\tau_{\mu}
\label{ksemu20}
\end{eqnarray}
where $\psi$ and $\tau_{\mu}$ are Majorana spinors depending only on the co-ordinates of $M_{11-n}$. Substituting \eqref{ksemu18} and \eqref{ksemu20} into \eqref{epsilon}, we get \eqref{ksemu21}. \\
\indent
Let us now analyze the integrability conditions of \eqref{ksem9}, that is
\begin{eqnarray}
[\widetilde{\nabla}_a,\widetilde{\nabla}_b]\psi=\frac{1}{4}\widetilde{R}_{abcd}\Gamma^{cd}\psi~.
\label{integrab11}
\end{eqnarray}
Using \eqref{ksem9} and \eqref{sigman2}, equation \eqref{integrab11} implies\footnote{Notice that $\Gamma_{a_1a_2\dots a_k}=0$ for $k\ge 7$, since $n\ge 5$.}
\begin{eqnarray}
\frac{1}{4}\widetilde{R}_{abcd}\Gamma^{cd}\psi&=&\bigg(-\frac{1}{144}\Gamma_{[a}^{~~c_1c_2c_3c_4}(\widetilde{\nabla}_{b]} X)_{c_1c_2c_3c_4}+\frac{1}{18}(\widetilde{\nabla}_{[a}X)_{b]c_1c_2c_3}\Gamma^{c_1c_2c_3}
\nonumber \\
&-&\frac{1}{864}X_a^{~~c_1c_2c_3}X_{bd_1d_2d_3}\Gamma_{c_1c_2c_3}^{~~~~~~~d_1d_2d_3}+\frac{1}{432}X^{c_1c_2c_3 c_4}X_{[a|c_1d_1d_2}\Gamma_{c_2c_3c_4|b]}^{~~~~~~~~~d_1d_2}
\nonumber \\
&+&\frac{1}{576}X^{d_1d_2c_1c_2}X_{d_1d_2c_3c_4}\Gamma_{abc_1c_2}^{~~~~~~~c_3c_4}-\frac{1}{1728}X^2\Gamma_{ab}
\nonumber \\
&+&\frac{1}{216}X^{c_1c_2c_3 c_4}X_{c_1c_2c_3[a}\Gamma_{|c_4|b]}+\frac{1}{48}X_a^{~~c_1c_2c_3}X_{bc_1c_2d}\Gamma_{c_3}^{~~d}
\nonumber \\
&&+\frac{1}{108}X_{abc_1d}X^{c_1c_2c_3 c_4}\Gamma_{c_2c_3c_4}^{~~~~~~~d}+\frac{1}{72}X_{[a|c_1c_2d}X^{c_1c_2c_3c_4}\Gamma_{|b]c_3c_4}^{~~~~~~~d}\bigg)\psi
\nonumber \\
\label{integrabcompl}
\end{eqnarray}
Contracting \eqref{integrabcompl} with $\Gamma^b$ and using the bosonic field equations \eqref{Ein1}-\eqref{maxwell} and the Bianchi identities \eqref{Bianchiid}, we obtain
\begin{eqnarray}
&&\bigg(-\frac{1}{2}\Gamma^b\widetilde{\nabla}_a\widetilde{\nabla}_b A+\frac{A}{288}(\widetilde{\nabla}_aX)_{b_1b_2b_3 b_4}\Gamma^{b_1b_2b_3 b_4}+\frac{1}{72}(\widetilde{\nabla}_cA)X^{b_1b_2b_3c}\Gamma_{b_1b_2b_3a}
\nonumber \\
&+&\frac{1}{12}(\widetilde{\nabla}_cA)X_a^{~cb_1b_2}\Gamma_{b_1b_2}-\frac{A}{864}X^{b_1\dots b_4}X_{ab_1c_3c_4}\Gamma_{b_2b_3b_4}^{~~~~~~~c_3c_4}
\nonumber \\
&-&\frac{A}{576}X^{b_1b_2b_3 b_4}X_{b_1b_2c_3c_4}\Gamma_{ab_3b_4}^{~~~~~~c_3c_4}
+\frac{A}{1728}\Gamma_a X^2-\frac{A}{432}X^{b_1b_2b_3 b_4}X_{b_1b_2b_3a}\Gamma_{b_4}
\nonumber \\
&+&\frac{A}{144}X_{ac_1c_2b_3}X^{c_1c_2c_3c_4}\Gamma_{c_3c_4}^{~~~~~b_3}\bigg)\psi=0 
\label{integrab1}
\end{eqnarray}
which coincides with \eqref{ksem92}. Moreover, by contracting \eqref{integrab1} with $\Gamma^a$ and implementing again the bosonic field equations \eqref{Ein1}-\eqref{maxwell} and the Bianchi identities  \eqref{Bianchiid}, we recover \eqref{integrsusp1}. Hence, it follows that equations \eqref{ksemu26} and \eqref{ksem92} are integrability conditions of \eqref{ksem9}.

\appendix{$dS_6$ backgrounds: linear system}

The linear system associated to the KSE \eqref{eqm4}, with $\Psi$ given by \eqref{PSIPSI}, is 
\begin{eqnarray}
\sin\theta \textbf{e}_7\theta=0
\label{i=7first}
\end{eqnarray}
\begin{eqnarray}
\cos\theta \textbf{e}_7\theta=0
\end{eqnarray}
\begin{eqnarray}
\cos\theta\dbtilde{\Omega}_{7,78}+\sin\theta\dbtilde{\Omega}_{7,9\sharp}=0
\end{eqnarray}
\begin{eqnarray}
\sin\theta\dbtilde{\Omega}_{7,78}+\cos\theta\dbtilde{\Omega}_{7,9\sharp}=0
\end{eqnarray}
\begin{eqnarray}
\cos\theta\dbtilde{\Omega}_{7,79}-\sin\theta\dbtilde{\Omega}_{7,8\sharp}=0
\end{eqnarray}
\begin{eqnarray}
\sin\theta\dbtilde{\Omega}_{7,79}-\cos\theta\dbtilde{\Omega}_{7,8\sharp}=0
\end{eqnarray}
\begin{eqnarray}
\cos\theta\dbtilde{\Omega}_{7,89}+\sin\theta\dbtilde{\Omega}_{7,7\sharp}-2c\cos\theta=0
\end{eqnarray}
\begin{eqnarray}
\sin\theta\dbtilde{\Omega}_{7,89}+\cos\theta\dbtilde{\Omega}_{7,7\sharp}+2c\sin\theta=0
\end{eqnarray}
\begin{eqnarray}
\sin\theta \textbf{e}_8\theta=0
\end{eqnarray}
\begin{eqnarray}
\cos\theta \textbf{e}_8\theta=0
\end{eqnarray}
\begin{eqnarray}
\cos\theta\dbtilde{\Omega}_{8,78}+\sin\theta\dbtilde{\Omega}_{8,9\sharp}=0
\end{eqnarray}
\begin{eqnarray}
\sin\theta\dbtilde{\Omega}_{8,78}+\cos\theta\dbtilde{\Omega}_{8,9\sharp}=0
\end{eqnarray}
\begin{eqnarray}
\cos\theta\dbtilde{\Omega}_{8,79}-\sin\theta\dbtilde{\Omega}_{8,8\sharp}+2c\cos\theta=0
\end{eqnarray}
\begin{eqnarray}
\sin\theta\dbtilde{\Omega}_{8,79}-\cos\theta\dbtilde{\Omega}_{8,8\sharp}-2c\sin\theta=0
\end{eqnarray}
\begin{eqnarray}
\cos\theta\dbtilde{\Omega}_{8,89}+\sin\theta\dbtilde{\Omega}_{8,7\sharp}=0
\end{eqnarray}
\begin{eqnarray}
\sin\theta\dbtilde{\Omega}_{8,89}+\cos\theta\dbtilde{\Omega}_{8,7\sharp}=0
\end{eqnarray}
\begin{eqnarray}
\sin\theta \textbf{e}_9\theta=0
\end{eqnarray}
\begin{eqnarray}
\cos\theta \textbf{e}_9\theta=0
\end{eqnarray}
\begin{eqnarray}
\cos\theta\dbtilde{\Omega}_{9,78}+\sin\theta\dbtilde{\Omega}_{9,9\sharp}-2c\cos\theta=0
\label{eq366}
\end{eqnarray}
\begin{eqnarray}
\sin\theta\dbtilde{\Omega}_{9,78}+\cos\theta\dbtilde{\Omega}_{9,9\sharp}+2c\sin\theta=0
\label{eq377}
\end{eqnarray}
\begin{eqnarray}
\cos\theta\dbtilde{\Omega}_{9,79}-\sin\theta\dbtilde{\Omega}_{9,8\sharp}=0
\end{eqnarray}
\begin{eqnarray}
\sin\theta\dbtilde{\Omega}_{9,79}-\cos\theta\dbtilde{\Omega}_{9,8\sharp}=0
\end{eqnarray}
\begin{eqnarray}
\cos\theta\dbtilde{\Omega}_{9,89}+\sin\theta\dbtilde{\Omega}_{9,7\sharp}=0
\end{eqnarray}
\begin{eqnarray}
\sin\theta\dbtilde{\Omega}_{9,89}+\cos\theta\dbtilde{\Omega}_{9,7\sharp}=0
\end{eqnarray}
\begin{eqnarray}
-\sin\theta \textbf{e}_{\sharp}\theta+c\sin\theta=0
\label{eq34}
\end{eqnarray}
\begin{eqnarray}
\cos\theta \textbf{e}_{\sharp}\theta-c\cos\theta=0
\label{eq35}
\end{eqnarray}
\begin{eqnarray}
\cos\theta\dbtilde{\Omega}_{\sharp,78}+\sin\theta\dbtilde{\Omega}_{\sharp,9\sharp}=0
\end{eqnarray}
\begin{eqnarray}
\sin\theta\dbtilde{\Omega}_{\sharp,78}+\cos\theta\dbtilde{\Omega}_{\sharp,9\sharp}=0
\end{eqnarray}
\begin{eqnarray}
\cos\theta\dbtilde{\Omega}_{\sharp,79}-\sin\theta\dbtilde{\Omega}_{\sharp,8\sharp}=0
\end{eqnarray}
\begin{eqnarray}
\sin\theta\dbtilde{\Omega}_{\sharp,79}-\cos\theta\dbtilde{\Omega}_{\sharp,8\sharp}=0
\end{eqnarray}
\begin{eqnarray}
\cos\theta\dbtilde{\Omega}_{\sharp,89}+\sin\theta\dbtilde{\Omega}_{\sharp,7\sharp}=0
\end{eqnarray}
\begin{eqnarray}
\sin\theta\dbtilde{\Omega}_{\sharp,89}+\cos\theta\dbtilde{\Omega}_{\sharp,7\sharp}=0~.
\label{i=sharplast}
\end{eqnarray}

\appendix{$dS_5$ backgrounds: linear system}

The linear system associated to the KSE \eqref{main1}, with $\Psi$ given by \eqref{psithisstage3}, reads (here we split the $\mathfrak{su}(3)$ indices as $\{1, P\}$ for $P=2,3$)
\begin{eqnarray}
\check{\textbf{e}}_1 g_1+\frac{1}{2}g_1\big(\check{\Omega}_{1,1\bar{1}}+\check{\Omega}_{1,P}^{~~~~P}\big)=0
\label{1star}
\end{eqnarray}
\begin{eqnarray}
\check{\textbf{e}}_1 g_1-\frac{1}{2}g_1\big(\check{\Omega}_{1,1\bar{1}}+\check{\Omega}_{1,P}^{~~~~P}\big)+\frac{i\sqrt{2}}{4}g_2A^{2}\check{G}_{1\bar{1}}+i\sqrt{2}g_2\sqrt{k}A^{-3}=0
\nonumber \\
\label{2star}
\end{eqnarray}
\begin{eqnarray}
&&\check{\textbf{e}}_1g_2+\frac{1}{2}\bar{g}_3\check{\Omega}_{1,PQ}\epsilon^{PQ}-\frac{1}{2}g_2\big(\check{\Omega}_{1,1\bar{1}}-\check{\Omega}_{1,P}^{~~~~P}\big)
\nonumber \\
&&+\frac{i\sqrt{2}}{4}g_1A^{2}\check{G}_{1\bar{1}}+i\sqrt{2}g_1\sqrt{k}A^{-3}=0
\label{3star}
\end{eqnarray}
\begin{eqnarray}
\check{\textbf{e}}_1g_2+\frac{1}{2}g_3\check{\Omega}_{1,\bar{P}\bar{Q}}\epsilon^{\bar{P}\bar{Q}}+\frac{1}{2}g_2\big(\check{\Omega}_{1,1\bar{1}}-\check{\Omega}_{1,P}^{~~~~P}\big)=0
\label{4star}
\end{eqnarray}
\begin{eqnarray}
\check{\textbf{e}}_1g_3-\frac{1}{2}g_2\check{\Omega}_{1,PQ}\epsilon^{PQ}+\frac{1}{2}g_3\big(\check{\Omega}_{1,1\bar{1}}+\check{\Omega}_{1,P}^{~~~~P}\big)=0
\label{5star}
\end{eqnarray}
\begin{eqnarray}
-\check{\textbf{e}}_1\bar{g}_3+\frac{1}{2}g_2\check{\Omega}_{1,\bar{P}\bar{Q}}\epsilon^{\bar{P}\bar{Q}}+\frac{1}{2}\bar{g}_3\big(\check{\Omega}_{1,1\bar{1}}+\check{\Omega}_{1,P}^{~~~~P}\big)=0
\label{6star}
\end{eqnarray}
\begin{eqnarray}
g_1\check{\Omega}_{1,\bar{P}\bar{Q}}\epsilon^{\bar{P}\bar{Q}}=0
\label{7star}
\end{eqnarray}
\begin{eqnarray}
-\frac{1}{2}g_1\check{\Omega}_{1,PQ}\epsilon^{PQ}+\frac{i\sqrt{2}}{4}g_3A^{2}\check{G}_{1\bar{1}}+i\sqrt{2}g_3\sqrt{k}A^{-3}=0
\label{8star}
\end{eqnarray}
\begin{eqnarray}
g_1\check{\Omega}_{1,\bar{1}}^{~~~~P}+\frac{i\sqrt{2}}{4}g_2A^{2}\check{G}_1^{~~P}-\frac{i\sqrt{2}}{4}\bar{g}_3A^{2}\check{G}_{1Q}\epsilon^{QP}=0
\label{9star}
\end{eqnarray}
\begin{eqnarray}
g_3\check{\Omega}_{1,\bar{1}}^{~~~~P}+g_2\check{\Omega}_{1,\bar{1}Q}\epsilon^{QP}+\frac{i\sqrt{2}}{4}g_1A^{2}\check{G}_{1Q}\epsilon^{QP}=0
\label{10star}
\end{eqnarray}
\begin{eqnarray}
-g_1\check{\Omega}_{1,1Q}\epsilon^{QP}+\frac{i\sqrt{2}}{4}g_3A^{2}\check{G}_1^{~~P}+\frac{i\sqrt{2}}{4}g_2A^{2}\check{G}_{1Q}\epsilon^{QP}=0
\label{11star}
\end{eqnarray}
\begin{eqnarray}
\bar{g}_3\check{\Omega}_{1,1Q}\epsilon^{QP}-g_2\check{\Omega}_{1,1}^{~~~~P}+\frac{i\sqrt{2}}{4}g_1A^{2}\check{G}_1^{~~P}=0
\label{12star}
\end{eqnarray}
\begin{eqnarray}
\check{\textbf{e}}_P g_1+\frac{1}{2}g_1\big(\check{\Omega}_{P,1\bar{1}}+\check{\Omega}_{P,Q}^{~~~~Q}\big)-\frac{i\sqrt{2}}{4}g_2A^2\check{G}_{P1}=0
\label{1starV}
\end{eqnarray}
\begin{eqnarray}
\check{\textbf{e}}_P g_1 -\frac{1}{2}g_1\big(\check{\Omega}_{P,1\bar{1}}+\check{\Omega}_{P,Q}^{~~~~Q}\big)+\frac{i\sqrt{2}}{4}g_2A^2\check{G}_{P\bar{1}}=0
\label{2starV}
\end{eqnarray}
\begin{eqnarray}
\check{\textbf{e}}_P g_2+\frac{1}{2}\bar{g}_3 \check{\Omega}_{P,MN}\epsilon^{MN}-\frac{1}{2}g_2\big(\check{\Omega}_{P,1\bar{1}}-\check{\Omega}_{P,Q}^{~~~~Q}\big)+\frac{i\sqrt{2}}{4}g_1A^2\check{G}_{P\bar{1}}=0
\nonumber \\
\label{3starV}
\end{eqnarray}
\begin{eqnarray}
\check{\textbf{e}}_P g_2+\frac{1}{2} g_3 \check{\Omega}_{P,\bar{M}\bar{N}}\epsilon^{\bar{M}\bar{N}}+\frac{1}{2}g_2\big(\check{\Omega}_{P,1\bar{1}}-\check{\Omega}_{P,Q}^{~~~~Q}\big)-\frac{i\sqrt{2}}{4}g_1A^2\check{G}_{P1}=0
\nonumber \\
\label{4starV}
\end{eqnarray}
\begin{eqnarray}
\check{\textbf{e}}_P g_3-\frac{1}{2}g_2\check{\Omega}_{P,MN}\epsilon^{MN}+\frac{1}{2}g_3\big(\check{\Omega}_{P,1\bar{1}}+\check{\Omega}_{P,Q}^{~~~~Q}\big)=0
\label{5starV}
\end{eqnarray}
\begin{eqnarray}
-\check{\textbf{e}}_P\bar{g}_3+\frac{1}{2}g_2\check{\Omega}_{P,\bar{M}\bar{N}}\epsilon^{\bar{M}\bar{N}}+\frac{1}{2}\bar{g}_3\big(\check{\Omega}_{P,1\bar{1}}+\check{\Omega}_{P,Q}^{~~~~Q}\big)=0
\label{6starV}
\end{eqnarray}
\begin{eqnarray}
\frac{1}{2}g_1\check{\Omega}_{P,\bar{M}\bar{N}}\epsilon^{\bar{M}\bar{N}}+\frac{i\sqrt{2}}{4}\bar{g}_3A^2\check{G}_{P1}=0
\label{7starV}
\end{eqnarray}
\begin{eqnarray}
-\frac{1}{2}g_1\check{\Omega}_{P,MN}\epsilon^{MN}+\frac{i\sqrt{2}}{4}g_3A^2\check{G}_{P\bar{1}}=0
\label{8starV}
\end{eqnarray}
\begin{eqnarray}
g_1\check{\Omega}_{P,\bar{1}}^{~~~~Q}+\frac{i\sqrt{2}}{4}g_2A^2\check{G}_P^{~~Q}-\frac{i\sqrt{2}}{4}\bar{g}_3A^2\check{G}_{PM}\epsilon^{MQ}+\delta_P^{~~Q}i\sqrt{2}g_2\sqrt{k}A^{-3}=0
\nonumber \\
\label{9starV}
\end{eqnarray}
\begin{eqnarray}
g_3\check{\Omega}_{P,\bar{1}}^{~~~~Q}+g_2\check{\Omega}_{P,\bar{1}M}\epsilon^{MQ}+\frac{i\sqrt{2}}{4}g_1A^2\check{G}_{PM}\epsilon^{MQ}=0
\label{10starV}
\end{eqnarray}
\begin{eqnarray}
-g_1\check{\Omega}_{P,1M}\epsilon^{MQ}+\frac{i\sqrt{2}}{4}g_3A^2\check{G}_P^{~~Q}+\frac{i\sqrt{2}}{4}g_2A^2\check{G}_{PM}\epsilon^{MQ}+\delta_P^{~~Q}i\sqrt{2}g_3\sqrt{k}A^{-3}=0
\nonumber \\
\label{11starV}
\end{eqnarray}
\begin{eqnarray}
\bar{g}_3\check{\Omega}_{P,1M}\epsilon^{MQ}-g_2\check{\Omega}_{P,1}^{~~~~Q}+\frac{i\sqrt{2}}{4}g_1A^2 \check{G}_P^{~~Q}+\delta_P^{~~Q}i\sqrt{2}g_1\sqrt{k}A^{-3}=0~.
\nonumber \\
\label{12starV}
\end{eqnarray}
Moreover, the linear system associated to the algebraic condition \eqref{main2}, with $\Psi$ given by \eqref{psithisstage3}, is given by 
\begin{eqnarray}
\frac{\sqrt{k}}{2}g_1+\frac{A^{5}}{12}g_1(\check{G}_{1\bar{1}}+\check{G}_{P}^{~~P})-\frac{i\sqrt{2}}{2}g_2A^{2}\check{\nabla}_1A=0
\label{dopoMajalg1}
\end{eqnarray}
\begin{eqnarray}
\frac{\sqrt{k}}{2}g_3-\frac{A^{5}}{12}g_2\check{G}_{PQ}\epsilon^{PQ}+\frac{A^{5}}{12}g_{3}(\check{G}_{1\bar{1}}+\check{G}_{P}^{~~P})=0
\label{dopoMajalg2}
\end{eqnarray}
\begin{eqnarray}
\frac{\sqrt{k}}{2}g_2+\frac{A^{5}}{12}\bar{g}_3\check{G}_{PQ}\epsilon^{PQ}-\frac{A^{5}}{12}g_2(\check{G}_{1\bar{1}}-\check{G}_{P}^{~~P})+\frac{i\sqrt{2}}{2}A^{2}g_1\check{\nabla}_{\bar{1}}A=0
\nonumber \\
\label{dopoMajalg3}
\end{eqnarray}
\begin{eqnarray}
\frac{i\sqrt{2}}{2}A^{2}g_3\check{\nabla}_{\bar{1}}A-\frac{A^{5}}{12}g_1\check{G}_{PQ}\epsilon^{PQ}=0
\label{dopoMajalg4}
\end{eqnarray}
\begin{eqnarray}
\frac{i\sqrt{2}}{2}g_1A^{2}\check{\nabla}^PA+\frac{A^{5}}{6}\bar{g}_3\check{G}_{1Q}\epsilon^{QP}-\frac{A^{5}}{6}g_2\check{G}_1^{~~P}=0
\label{dopoMajalg5}
\end{eqnarray}
\begin{eqnarray}
\frac{i\sqrt{2}}{2}A^{2}g_2\check{\nabla}^PA-\frac{i\sqrt{2}}{2}A^{2}\bar{g}_3\check{\nabla}_QA\epsilon^{QP}+\frac{A^{5}}{6}g_1 \check{G}_{\bar{1}}^{~~P}=0~.
\label{dopoMajalg6}
\end{eqnarray}

\appendix{Solutions with parallel form}

In this Appendix, we provide some details about the derivation of \eqref{fszizi}-\eqref{usefulhij}. First of all, using \eqref{Gcoordinate}, equation \eqref{constG} implies
\begin{eqnarray}
f^2\bigg(\bigg(\frac{\partial f}{\partial s}\bigg)^2+h^{ij}\frac{\partial f}{\partial z^i}\frac{\partial f}{\partial z^j}\bigg)=c^2
\label{constantC}
\end{eqnarray}
where $c$ is constant. Inserting \eqref{constantC} into \eqref{pde1}, we get
\begin{eqnarray}
f^3\frac{\partial^2 f}{\partial s^2}+3f^2\bigg(\frac{\partial f}{\partial s}\bigg)^2-2c^2=0~.
\label{pdesdep}
\end{eqnarray}
Defining $f=B^{\frac{1}{4}}$, equation \eqref{pdesdep} implies
\begin{eqnarray}
\frac{\partial^2B}{\partial s^2}=8c
\end{eqnarray}
whose solution is given by
\begin{eqnarray}
B=4cs^2-4 g_1(z) s+4g_2(z)
\end{eqnarray}
where $g_1(z)$ and $g_2(z)$ are arbitrary functions. Hence
\begin{eqnarray}
f(s,z^i)=\bigg(4c^2 s^2-4 g_1(z) s+4 g_2(z)\bigg)^{\frac{1}{4}}~.
\label{fszi}
\end{eqnarray}
Inserting \eqref{fszi} into \eqref{pde2}, we get
\begin{eqnarray}
\frac{\partial g_1(z)}{\partial z^i}=0
\end{eqnarray}
hence
\begin{eqnarray}
g_1=q
\label{g1zconst}
\end{eqnarray}
for constant $q$. Inserting \eqref{g1zconst} in \eqref{fszi}, we get
\begin{eqnarray}
f(s,z^i)=\bigg(4c^2 s^2-4 q s+4 P(z)\bigg)^{\frac{1}{4}}~.
\label{fsz2}
\end{eqnarray}
Substituting \eqref{fsz2} back into \eqref{pdesdep}, we get
\begin{eqnarray}
(\ciD P)^2=4c^2P-q^2~.
\label{EQforP2}
\end{eqnarray}
Also, substituting \eqref{fsz2} into \eqref{pde3}, we obtain
\begin{eqnarray}
\ciD_i\ciD_j P=2c^2 h_{ij}~.
\label{EQforP}
\end{eqnarray}
Defining
\begin{eqnarray}
s^{'}=s-\frac{q}{2c^2}~,~~~~P^{'}(z)=P(z)-\frac{q^2}{4c^2}
\end{eqnarray}
and dropping the primes, equations \eqref{fsz2}, \eqref{EQforP2} and \eqref{EQforP} imply \eqref{fszizi}-\eqref{usefulhij}, respectively.

\section*{Acknowledgments}

DF is partially supported by the STFC DTP Grant ST/S505742.

\section*{Data Management}

No additional research data beyond the data presented and cited in this work are needed to validate the research findings in this work.

\end{document}